# Report of the Working Group on Precision Measurements

**Conveners**: Raymond Brock[a], Jens Erler[b], Young-Kee Kim[c], and William Marciano[d]
**Working Group Members**: William Ashmanskas[e], Ulrich Baur[f], John Ellison[g], Mark Lancaster[h], Larry Nodulman[i], John Rha[j], David Waters[k], John Womersley[l]

[a]Michigan State University, East Lansing, MI 48824

[b]University of Pennsylvania, Philadelphia, PA 19104

[c]University of California, Berkeley, CA 94720

[d]Brookhaven National Laboratory, Upton, NY 11973

[e]University of Chicago, Chicago, IL 60637

[f]State University of New York, Buffalo, NY 14260

[g]University of California, Riverside, CA 92521

[h]University College, London WC1E 6BT, U.K.

[i]Argonne National Laboratory, Argonne, IL 60439

[j]University of California, Riverside, CA 92521

[k]Oxford University, Oxford, OX1 3RH, U.K.

[l]Fermilab, Batavia, IL 60510

## Overview

Precision measurements of electroweak quantities are carried out to test the Standard Model (SM). In particular, measurements of the top quark mass, $m_{top}$, when combined with precise measurements of the $W$ mass, $M_W$, and the weak mixing angle, $\sin^2 \bar{\theta}_W$, make it possible to derive indirect constraints on the Higgs boson mass, $M_H$, via top quark and Higgs boson electroweak radiative corrections to $M_W$. Comparison of these constraints on $M_H$ with the mass obtained from direct observation of the Higgs boson in future collider experiments will be an important test of the SM.

In this report, the prospects for measuring the $W$ parameters (mass and width) and the weak mixing angle in Run II are discussed, and a program for extracting the probability distribution function of $M_H$ is described. This is done in the form of three largely separate contributions.

The first contribution describes in detail the strategies of measuring $M_W$ and the $W$ width, $\Gamma_W$, at hadron colliders, and discusses the statistical, theoretical and detector specific uncertainties expected in Run II. The understanding of electroweak radiative corrections is crucial for precision $W$ mass measurements. Recently, improved calculations of the electroweak radiative corrections to $W$ and $Z$ boson production in hadronic collisions became available. These calculations are summarized and preliminary results from converting the theoretical weighted Monte Carlo program into an event generator are described. The traditional method of extracting $M_W$ from the lineshape of the transverse mass distribution has been the optimal technique for the extraction of $M_W$ in the low luminosity environment of Run I. Other techniques may cancel some of the systematic and statistical uncertainties resulting in more precise measurements for the high luminosities expected in Run II. Measuring the $W$ mass from fits of the transverse momentum distributions of the $W$ decay products and the ratio of the transverse masses of the $W$ and $Z$ bosons are investigated in some detail. Finally, the precision expected for the $W$ mass in Run II is compared with that from current LEP II data, and the accuracy one might hope to achieve at the LHC and a future linear $e^+e^-$ collider.

In the second contribution, a study of the measurement of the forward-backward asymmetry, $A_{FB}$, in $e^+e^-$ and $\mu^+\mu^-$ events is presented. The forward-backward asymmetry of $\ell^+\ell^-$ events in Run II can yield a measurement of the effective weak mixing angle $\sin^2\bar\theta_{\rm W}$ and can provide a test of the standard model $\gamma^*/Z$ interference at $\ell^+\ell^-$ invariant masses well above the 200 GeV center of mass energy of the LEP collider. The asymmetry at large partonic center of mass energies can also be used to study the properties of possible new neutral gauge bosons, and to search for compositeness and large extra dimensions. Estimates of the statistical and systematic uncertainties expected in Run II for $A_{FB}$ and $\sin^2\bar\theta_{\rm W}$ are given. The uncertainty for $\sin^2\bar\theta_{\rm W}$ is compared with the precision expected from LHC experiments, and from a linear collider operating at the $Z$ pole.

The third contribution summarizes the features of the FORTRAN package GAPP which performs a fit to the electroweak observables and extracts the probability distribution function of $M_H$.

# Measurement of the $W$ Boson Mass and Width

William Ashmanskas[a], Ulrich Baur[b], Raymond Brock[c], Young-Kee Kim[d], Mark Lancaster[e], Larry Nodulman[f], David Waters[g], John Womersley[h]

[a]University of Chicago, Chicago, IL 60637

[b]State University of New York, Buffalo, NY 14260

[c]Michigan State University, East Lansing, MI 48824

[d]University of California, Berkeley, CA 94720

[e]University College, London WC1E 6BT, U.K.

[f]Argonne National Laboratory, Argonne, IL 60439

[g]Oxford University, Oxford, OX1 3RH, U.K.

[h]Fermilab, Batavia, IL 60510

We discuss the prospects for measuring the $W$ mass and width in Run II. The basic techniques used to measure $M_W$ are described and the statistical, theoretical and detector-related uncertainties are discussed in detail. Alternative methods of measuring the $W$ mass at the Tevatron and the prospects for $M_W$ measurements at other colliders are also described.

## 1. Introduction

Measuring the $W$ mass, $M_W$, and width, $\Gamma_W$ are important objectives for the Tevatron experiments in Run II. The goal for the $W$ mass measurement at the Tevatron in Run II is determined by three factors: the direct measurement of the LEP II experiments, the indirect determination from within the Standard Model (SM), and the ultimate precision on the measured top quark mass. The expectations for LEP II appear to be an overall uncertainty of approximately $\pm 35$ MeV/c$^2$ [1]. The indirect determination is at the $\pm 30$ MeV/c$^2$ level and is not likely to significantly improve given the end of the LEP and SLC programs. Finally, the top quark mass precision may reach the $\pm 2$ GeV/c$^2$ level, which corresponds to a parametric uncertainty of $\delta M_W = 12$ MeV/c$^2$ [2]. The constraint provided by a successful $\pm 30$ MeV/c$^2$ $W$ boson mass measurement per experiment per channel[1] would have an impact on electroweak global fitting comparable to that of the LEP $Z$ asymmetries. If the ultimate precision on the $M_W$ determination could reach $\sim \pm 30$ MeV/c$^2$, then the bound on the Higgs boson mass would reach approximately $\pm 30$ GeV/c$^2$ [3]. With the best fit central value close to the current LEP II direct search lower limit of $M_H > 113.2$ GeV/c$^2$ [4], considerable pressure can be brought to bear on the SM.

This document is structured as follows. The basic techniques used to measure the $W$ mass and width are briefly reviewed in section 2. The statistical and detector-related uncertainties affecting the $W$ mass and width measurements are discussed in more detail in section 3 and section 4, respectively. A number of systematic uncertainties clearly do not scale statistically and these are addressed separately in section 5. The expected errors on the measured $W$ mass in Run II using the conventional transverse mass method and the $W$ width are summarized in section 6. Alternative methods of measuring the $W$ mass at the Tevatron are described in section 7 and prospects for $M_W$ measurements at other colliders are discussed in section 8. Finally, some theoretical considerations important for future $W$ mass measurements are brought up in section 9. Section 10 concludes this document.

## 2. $M_W$ and $\Gamma_W$ Measurements from the $M_T$ Lineshape

The determination of $M_W$ depends on the two body nature of the $W$ decay: $W \to \ell\nu_\ell$. The kinematical Jacobian peak and sharp edge at the value of $M_W/2$ is easily observed in the measurement of the transverse momentum of either of the leptons. In practice, the situation is difficult due to both challenging experimental issues and the fact that phenomenological assumptions must be made in order to perform the analysis. Because the standard measurable cannot be written in closed form, an unbinned maximum likelihood calculation is required. Figure 1 shows a calculation of $p_T^e$ (unsmeared) with $p_T^W = 0$; the effect of finite $p_T^W$; and the inclusion of detector smearing effects. It is apparent that $p_T^e$ is very sensitive to the transverse motion of the $W$ boson.

---
[1]While the measurements from the different channels and different experiments provide cross checks, the combined measurement is not expected to yield a much better precision than a single measurement because of large common uncertainties.

Table 1
Advantages and disadvantages to an $M_W$ determination via transverse mass or lepton transverse momenta.

| Measurable | $p_T^W$ sensitivity | resolution sensitivity |
|---|---|---|
| $M_T$ | small | significant |
| $p_T^e$ | significant | small |
| $p_T^\nu$ | significant | significant |

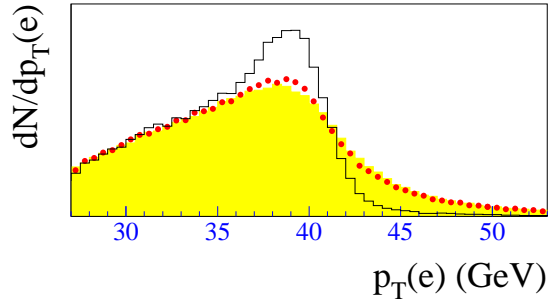

Figure 1. The effects of resolution and the finite $p_T^W$ on $p_T^e$ in $W$ boson decay. The histogram shows $p_T^W$ without detector smearing and for $p_T^W = 0$. The dots include the effects of adding finite $p_T^W$, while the shaded histogram includes the effects of detector resolutions. The effects are calculated for the DØ Run I detector resolutions.

Historically, precise understanding of $p_T^W$ has been lacking, although it is currently modeled by measurable parameters through the resummation formalism of Collins, Soper, and Sterman [ 5]. For this reason, the transverse mass quantity was suggested [ 6] and has been the traditional measurable. It is defined by

$$M_T = \sqrt{2p_T^\ell p_T^\nu (1 - \cos(\phi_{\ell,\nu}))}, \qquad (1)$$

where $\phi_{\ell,\nu}$ is the angle between the charged lepton and the neutrino in the transverse plane. The observables are the lepton transverse energy or momentum $\vec{p}_T^\ell$ and the non-lepton transverse energy $\vec{u}$ (recoil transverse energy against the $W$), from which the neutrino momentum $\vec{p}_T^\nu$ and the transverse mass $M_T$ are derived. Figure 2 shows that the sensitivity of $M_T$ to $p_T^W$ is nearly negligible. While considerably more stable to the phenomenology of the production model, the requirement that the neutrino direction be accurately measured leads to a set of experimental requirements which are difficult in practice to control. So, there are different benefits and challenges among the direct measurements of the transverse quantities, $p_T^\ell$, $p_T^\nu$, and $M_T$. Table 1 lists these relative pros and cons of the transverse mass and transverse momentum measurements.

Both CDF and DØ have determined the $W$ boson mass using the transverse mass approach. The individual measurements of both experiments are shown in Table 2 and the overall combined result is

$$M_W = 80.452 \pm 0.062 \text{ GeV/c}^2. \qquad (2)$$

The $W$ boson width is precisely predicted in terms of well-measured SM masses and coupling strengths:

$$\Gamma_W = \frac{G_F M_W^3}{6\sqrt{2}\pi}\left[3 + 6\left(1 + \frac{\alpha_s(M_W)}{\pi} + \mathcal{O}(\alpha_s^2)\right)\right]$$
$$\times (1 + \mathcal{O}(1\%))$$
$$= 2.093 \pm 0.002 \text{ GeV} \qquad (3)$$

where the uncertainty is dominated by the experimental $M_W$ precision [ 7, 8]. The mass and width of the $W$ boson connect both theoretically and experimentally, as $\Gamma_W$ has been extracted from a lineshape analysis using techniques developed for the $W$ mass measurement. Combining CDF electron and muon data from 1994–95 yields a result with 140 MeV precision [ 9]:

$$\Gamma_W = 2.04 \pm 0.11 \text{ (stat)} \pm 0.09 \text{ (syst) GeV}. \qquad (4)$$

In this measurement, $u < 20$ GeV is required to improve the $M_T$ resolution and to reduce backgrounds.

Figure 3 shows the dependence of the $M_T$ spectrum on $\Gamma_W$. In the region $M_T > 100$ GeV/c$^2$, the lineshape is sensitive to $\Gamma_W$ but relatively insensitive to uncertainties in the $p_T^\nu$ resolution. Thus, $\Gamma_W$ is extracted from a fit to the region 100 GeV/c$^2 < M_T < 200$ GeV/c$^2$, after signal and background templates are normalized to the data in the region 40 GeV/c$^2 < M_T < 200$ GeV/c$^2$. Figure 4 shows the fits to the CDF electron and muon data. The upper limit $M_T < 200$ GeV/c$^2$ is somewhat arbitrary.

The measurement of $\Gamma_W$ depends on a precise determination of the transverse mass lineshape. Thus, the same error sources contribute to both the $W$ mass and width measurement. In the following we discuss these sources, concentrating on how they impact the $W$ mass measurement. Run II projections for the individual uncertainties contributing to the $W$ width measurement are presented in section 6.

Table 2
Tevatron results for $M_W$. $N_W$ is the number of $W$ boson events observed. Scale and non-scale systematic errors are listed separately.

| Experiment | $\int \mathcal{L} dt$ pb$^{-1}$ | $N_W$ | $M_W$ GeV/c$^2$ | ± stat GeV/c$^2$ | ± sys GeV/c$^2$ | ± scale GeV/c$^2$ |
|---|---|---|---|---|---|---|
| CDF Run 0 e | 4.4 | 1130 | 79.91 | 0.35 | 0.24 | 0.19 |
| CDF Run 0 $\mu$ | 4.4 | 592 | 79.90 | 0.53 | 0.32 | 0.08 |
| CDF Run Ia e | 18.2 | 5718 | 80.490 | 0.145 | 0.130 | 0.120 |
| DØ Run Ia e | 12.8 | 5982 | 80.350 | 0.140 | 0.165 | 0.160 |
| CDF Run Ia $\mu$ | 19.7 | 3268 | 80.310 | 0.205 | 0.120 | 0.050 |
| CDF Run Ib e | 84 | 30,100 | 80.473 | 0.065 | 0.054 | 0.075 |
| DØ Run Ib e | 82 | 28,323 | 80.440 | 0.070 | 0.070 | 0.065 |
| DØ Run Ib e, forward | 82 | 11,089 | 80.757 | 0.107 | 0.091 | 0.181 |
| CDF Run Ib $\mu$ | 80 | 14,700 | 80.465 | 0.100 | 0.057 | 0.085 |

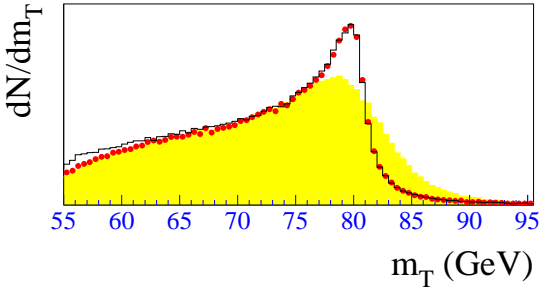

Figure 2. The effects of resolution and the finite $p_T^W$ on $M_T$ in $W \to e\nu$. The histogram shows $M_T$ without detector smearing and for $p_T^W = 0$. The dots include the effects of adding finite $p_T^W$, while the shaded histogram includes the effects of detector resolutions. The effects are calculated for the DØ Run I detector resolutions.

### 3. Statistical Uncertainties in the $M_W$ Determination

In order to reach the target precision for $M_W$, considerable luminosity will be required. Presuming that Run II is to deliver an integrated luminosity of 2 fb$^{-1}$, the statistical precision on $M_W$ can be estimated from the existing data (see Table 2). Figure 5 shows the $W$ statistical uncertainties in these measurements as a function of $1/\sqrt{N_W}$, demonstrating a predictable extrapolation to $N_W \simeq 700,000$ which corresponds to a Run II dataset per experiment per channel. The statistical uncertainty from this extrapolation is approximately 13 MeV/c$^2$. For a goal of ±30 MeV/c$^2$ overall uncertainty, this leaves 27 MeV/c$^2$ available in the error budget which must be accounted for by all systematic uncertainties.

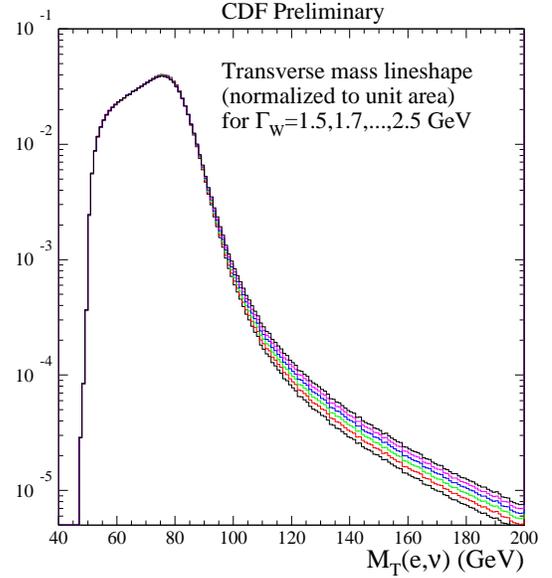

Figure 3. Dependence of the $M_T$ spectrum on $\Gamma_W$. Each spectrum is normalized to unit area.

### 4. Detector-specific Uncertainties in the $M_W$ Determination

After the lepton energy and momentum scales, the modeling of the $W$ recoil provided the largest systematic uncertainty in the CDF Run Ib $W$ mass measurement. Since $Z$ statistics dominates this number, it can be expected to be reduced significantly in Run II. Non-$Z$ related recoil systematics were estimated to enter at the 10 MeV/c$^2$ level, which is probably indicative of the limiting size of this error. The increase in the average number of overlapping minimum bias events in Run II may seriously impact the recoil model systematics, although various detector improvements may partly compensate for this.

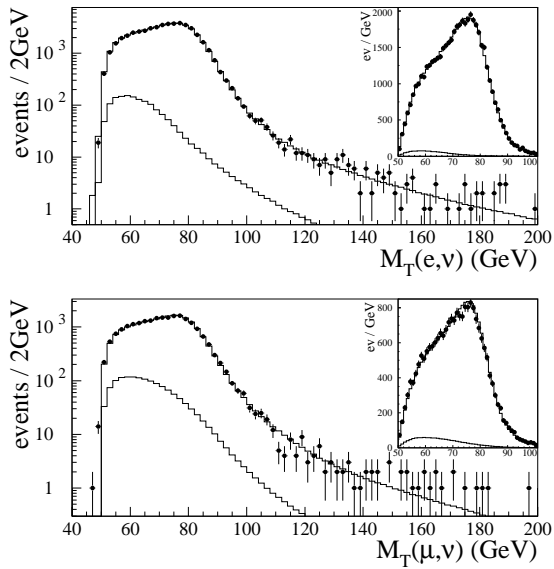
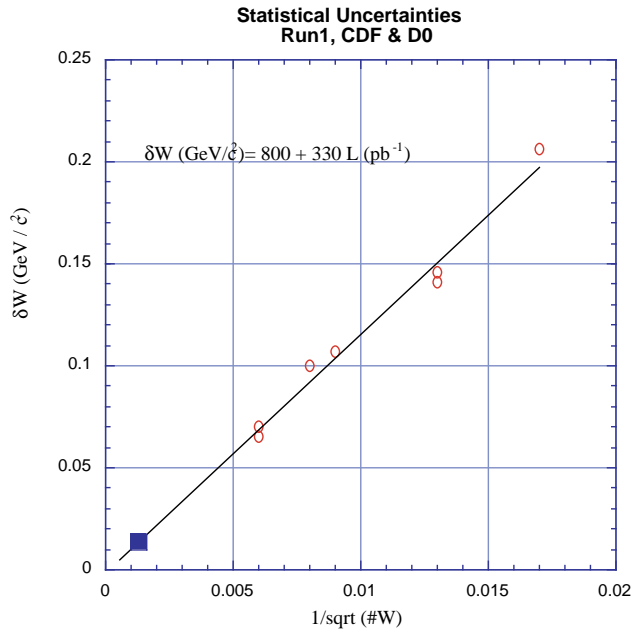

Figure 4. CDF 1994-95 $e$ and $\mu$ data, on a semilogarithmic scale, with best fits for $\Gamma_W$. The background estimates are also shown. The insets show the Jacobian peak regions on a linear scale.

Figure 5. Statistical uncertainties in Run I $M_W$ measurements. Each circle represents either a CDF or DØ measurement. The result of a straight line fit is shown. The shaded box is the approximate extrapolation to a 2 fb$^{-1}$ Run II result.

Much of the understanding of experimental systematics comes from a detailed study of the $Z$ bosons and hence as luminosity improves, systematic uncertainties should diminish in kind. Certainly, the scale and resolution of the recoil energy against the $W$ come from measurements of the $Z$ system. Likewise, background determination, underlying event studies, and selection biases depend critically, but not exclusively, on $Z$ boson data. Most importantly, the lepton energy and momentum scales depend solely on the $Z$ boson datasets.

Figure 6 shows the CDF and DØ systematic uncertainties for both electrons and muons as a function of $1/\sqrt{N_W}$ and in particular the calorimeter scale uncertainties for electrons. This latter important energy scale determination is currently tied to the determination of fiducial di-lepton decay resonances, notably the $Z$ boson, but also the $J/\psi$, $\Upsilon$ and the $E/p$ dependence on the energy $E$, using electrons from $W$ and $Z$ decays. As the statistical precision improves, the dominant feature of the scale determination becomes its value in the region of $M_W$,
so offsets and any low energy nonlinearities become relatively less important and hence reliance on the low mass resonances is reduced. On the other hand, for the muon momentum scale determination, where the observable is the curvature, low mass resonances are also important. Figure 6 suggests that this uncertainty is truly statistical in nature and extrapolates to approximately the 15 MeV/c$^2$ level. The ability to bound non-linearities using collider data may become a limiting source of error in Run II. Hence, the remaining systematic uncertainties must be controlled to a level of approximately 22 MeV/c$^2$ in order to reach the overall goal of $\pm 30$ MeV/c$^2$.

Figure 6 also shows the non-scale systematic uncertainties from both the CDF and DØ electron measurements of $M_W$ and the CDF muon measurement. Here the extrapolation is not as straightforward, but there is clearly a distinct statistical nature to these errors. That they appear to extrapolate to negative values suggests that the systematic uncertainties may contain a statistically independent component for both the muon and the electron analyses.

For both $M_W$ and $\Gamma_W$ analyses, the $Z \to \ell\ell$ data constrain both the lepton scales and resolutions and an empirical model of the hadronic recoil measurement. QED corrections are an issue in measuring the $Z$ mass, and the discussion of these corrections should be in terms of the $W/Z$ mass ratio. In a high-precision width measurement, more effort will also be needed to place bounds on possible tails in the lepton and recoil resolution functions. Uncertainty in the recoil measurement is predominantly statistical in how well model parameters are determined. Several cross checks which improve with statistics independently ensure the efficacy of the model.

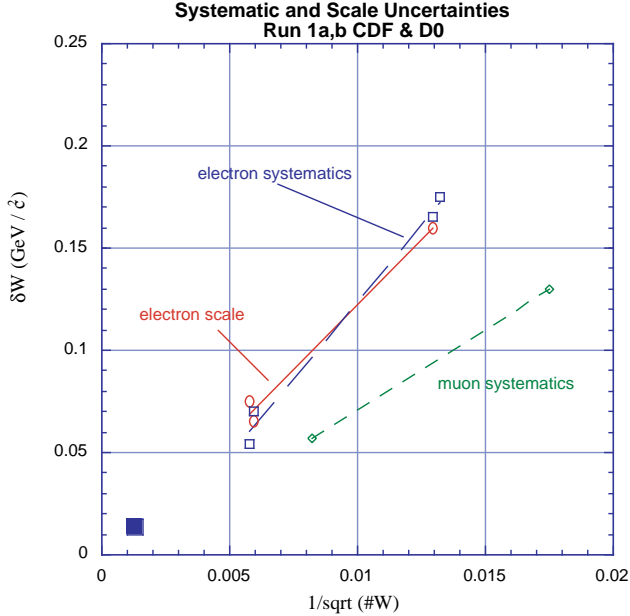

Figure 6. Systematic uncertainties for each Run I $M_W$ measurement. The open squares are the four electron measurements from CDF and DØ, the circles are the scale uncertainties from two DØ electron measurements and the Run Ib CDF measurement, and the diamonds are the systematic uncertainties (excluding scale) for the CDF muon measurements. The large box is the position of the extrapolated statistical uncertainties to the Run II luminosity. The lines are linear fits to each set of points.

Selection biases can be studied with various control samples, notably the second lepton originating from $Z$ decays. The QCD background can be also studied by varying cuts and studying control samples. The background from $W \to \tau\nu$ is well understood, and the background from $Z \to \ell\ell$ will be reduced for Run II since the tracking and muon coverages are improved for both experiments.

## 5. Theoretical Uncertainties in the $M_W$ Determination

The $M_T$ lineshape simulation requires a theoretical model, as a function of $\Gamma_W$ and $M_W$, of $\frac{d^3\sigma}{d\hat{s}\, dy\, dp_T}$, including correlations between $p_T$ and $\hat{s}$. For producing high-statistics fitting templates, a weighted Monte Carlo generator is useful, so that $M_W$, $\Gamma_W$, and the $p_T$ spectrum can be varied simply by reweighting events. Because the measurement of the recoil energy against the $W$, $\vec{u}$, is modeled empirically, the generator does not have to describe the recoil energy at the particle level. A detailed description of final-state QED radiation is important, because bremsstrahlung affects the isolation variables needed to select a clean $W$ sample.

The $W$ and $Z$ $p_T$ spectra are not calculable using perturbation theory at low $p_T$. In this region, the perturbative calculation must be augmented by a non-perturbative contribution which depends on three parameters (see section 5.2.1) which are tuned to fit the $Z \to \ell\ell$ data. Theoretical guidance is useful for choosing an appropriate set of parameters to vary. A strategy such as has been used in the CDF Run Ib analysis to use theory to extrapolate from the $Z$ $p_T$ distribution to the $W$ $p_T$ distribution seems to limit the effect of theoretical assumptions to $\pm 5$ MeV/c$^2$.

The parameters of parton distribution functions are also empirical, and seldom have quoted uncertainties. PDF uncertainties seem under control for Run I data but will need improvement to avoid becoming dominant in Run II. More work is needed to determine how both to minimize the impact of PDF uncertainties (e.g. by extending the lepton rapidity coverage of the measurements as done in the DØ analysis [ 10]) and to evaluate the effects of PDF uncertainties in precision measurements.

To date, *ad hoc* event generators have been used in the $W$ mass and width measurements. In Run II, these measurements will reach a precision of tens of MeV/c$^2$, requiring much more attention to detail in Monte Carlo calculations. Precision electroweak measurements in Run II should strive to use (possibly to develop) published, well documented Monte Carlo programs that are common to both collider experiments. In particular, the $M_W$ and $\Gamma_W$ measurements would benefit from a unified generator that incorporates state-of-the-art QED and electroweak calculations, uses a boson $p_T$ model tunable to Run II data, and correctly handles $W$ bosons that are produced far off-shell.

The $W$ width uncertainty in the $M_W$ measurement could become significant but assuming the SM $M_W$-$\Gamma_W$ relation, it won't be a dominant uncertainty.

## 5.1. Parton Distribution Functions

The transverse mass distribution is invariant under the longitudinal boost of the $W$ boson. However, the incomplete $\eta$ coverage of the detectors introduces a dependence of the measured $M_T$ distribution on the longitudinal momentum distribution of the produced $W$'s, determined by the PDF's. Therefore, quantifying the uncertainties in PDFs and the resulting uncertainties in the $W$ mass measurement is crucial.

### 5.1.1. Constraining PDFs from the Tevatron data

The measurement of the $W$ charge asymmetry at the Tevatron, which is sensitive to the ratio of $d$ to $u$ quark densities in the proton, is of direct benefit in constraining PDF effects in the $W$ mass measurement. This has been demonstrated by the CDF experiment. Following Ref. [ 11], they made parametric modifications to the MRS family of PDFs. These modifications with retuned parameters are listed in Table 3 and their predictions are compared to the $W$ lepton charge asymmetry measurement and the NMC $d/u$ data [ 12] in Fig. 7. From the variation among the six reference PDFs, an uncertainty of 15 MeV/$c^2$ was taken which is common to the electron and muon analyses.

Since the Run Ib charge asymmetry data is dominated by statistical uncertainties, we expect a smaller uncertainty for the Run II measurement. Measurements of Drell-Yan production at the Tevatron can be used to get further constraints on PDFs.

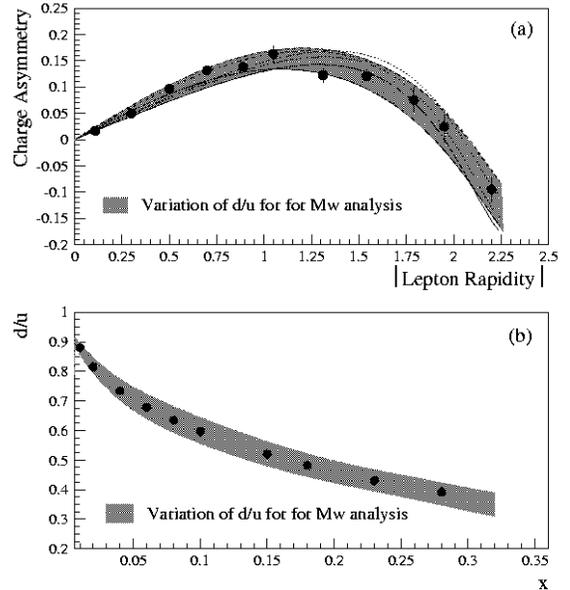

Figure 7. (a) The CDF measurement of the $W$ lepton charge asymmetry. (b) The NMC $d/u$ data evolved to $Q^2 = M_W^2$. The gray bands represent the range spanned by the six reference PDFs considered.

Table 3
Reference PDFs and modifications

| PDF set | Modification |
| --- | --- |
| MRS-T | $d/u \to d/u \times (1.07 - 0.07 e^{-8x})$ |
| MRS-R2 | $d/u \to d/u + 0.11 x \times (1 + x)$ |
| MRS-R1 | $d/u \to d/u \times (1.00 - 0.04 e^{-\frac{1}{2}(\frac{(x-0.07)}{0.015})^2})$ |

### 5.1.2. Reducing the PDF uncertainty with a larger $\eta$ coverage

Since the PDF uncertainty comes from the finite $\eta$ coverage of the detectors, it is expected to decrease with the more complete rapidity coverage of the Run II detectors. The advantage of a larger rapidity coverage has been demonstrated by the DØ experiment: the uncertainty on the $W$ mass measurement using their central calorimeter was 11 MeV/$c^2$, while that using both the central and end calorimeter was 7 MeV/$c^2$. With the upgraded calorimeters and trackers for the range $|\eta| > 1$, the CDF experiment can measure the $W$ mass over a larger rapidity range in Run II.

### 5.1.3. A Global Approach

There has been a systematic effort to map out the uncertainties allowed by available experimental constraints, both on the PDFs themselves and on physical observables derived from them. This approach will provide a more reliable estimate and may be the best course of action for precision measurements such as the $W$ mass or the $W$ production cross section. This has been emphasized at this workshop by the Parton Distributions Working Group [ 13].

## 5.2. $W$ Boson Transverse Momentum

The neutrino transverse momentum is estimated by combining the measured lepton transverse momentum and the $W$ recoil: $\vec{p}_T^{\,\nu} = -(\vec{p}_T^{\,l} + \vec{u})$. It is clear therefore that an understanding of both the underlying $W$ boson transverse momentum distribution and the corresponding detector response, usually called the recoil model, is crucial for a precision $W$ mass measurement. For the CDF Run Ib $W$ mass measurement, the systematic uncertainties from these two sources were estimated to be $15 - 20$ MeV/$c^2$ and $35 - 40$ MeV/$c^2$, respectively, in each channel [ 14].

### 5.2.1. Extracting the $p_T^W$ Distribution

The strategy employed in Run I, which is expected to be used also in Run II, is to extract the underlying $p_T^W$ distribution from the measured $p_T^Z$ distribution ($Y$ is the weak boson rapidity):

$$\frac{d^2\sigma}{dp_T^W dY} = \frac{d^2\sigma}{dp_T^Z dY} \times \frac{d^2\sigma/dp_T^W dY}{d^2\sigma/dp_T^Z dY} \;, \qquad (5)$$

where the ratio of the $W$ and $Z$ differential distributions is obtained from theory. This method relies on the fact that the observed $p_T^Z$ distribution suffers relatively little from detector smearing effects, allowing fits to be performed for the true distribution. The CDF Run Ib data and the results of a Monte Carlo simulation using the best fit parameters are compared in Fig. 8.

to $Z$ transverse momentum distributions contribute a further $\mathcal{O}(5)$ MeV/c$^2$ to the overall error. The two sources of uncertainty are compared for the CDF Run Ib $W \to \mu\nu$ analysis in Fig. 9.

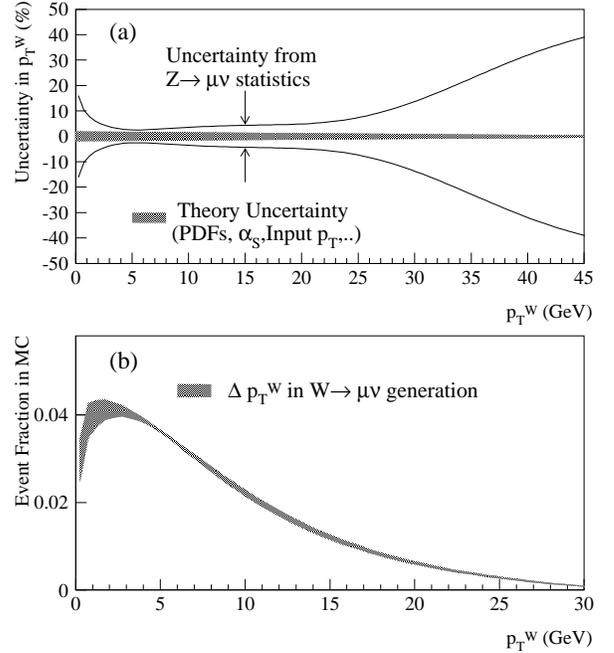

Figure 9. (a) A comparison of the two sources of uncertainty on the derived $p_T^W$ distribution. (b) The $p_T^W$ distribution extracted for the CDF Run Ib $W \to \mu\nu$ mass measurement.

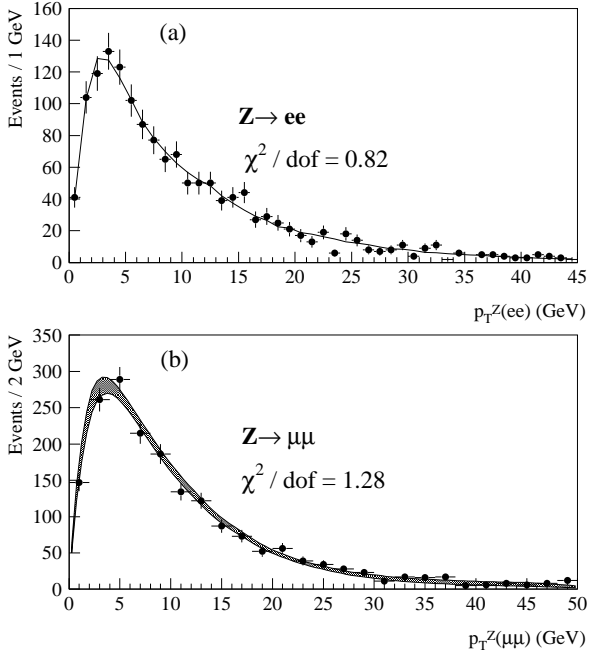

Figure 8. The observed $p_T^Z$ distributions in the electron and muon channels using the CDF Run Ib data. Also shown are the curves for the Monte Carlo simulation using the best fit parameters for the input $p_T^Z$ distribution.

The experimental uncertainties, as in many aspects of the $W$ mass measurement, are dominated by the available $Z \to \ell^+\ell^-$ statistics and should scale correspondingly with the delivered luminosity in Run II. Theoretical uncertainties in the ratio of $W$

The ratio of $W$ to $Z$ transverse momentum distributions used in Eq. (5) is taken from resummation calculations, which attempt to resum terms corresponding to multiple soft and collinear gluon emission to all orders. They thereby include the dominant contribution to the cross section at small boson $p_T$ that is missing in fixed order calculations. These perturbative calculations need to be augmented by a non-perturbative contribution which, in the case of impact-parameter ($b$) space resummations, is typically parameterized as a Sudakov form-factor with the following form:

$$\begin{aligned} F^{NP} = & \exp\left[-g_1 b^2 - g_2 b^2 \ln(Q/2Q_0) \right. \\ & \left. - g_1 g_3 b \ln(100 x_1 x_2)\right] \;, \end{aligned} \qquad (6)$$

where $Q_0$ is a low scale of $\mathcal{O}$(few) GeV and the parameters $g_1, g_2$ and $g_3$ must be obtained from fits to the data [15]. DØ has shown that the Run I $p_T^Z$ data is as sensitive to $g_1$ and $g_2$ as the low-energy Drell-Yan data that has largely been used to constrain these parameters in the past [ 16]. The Run II data will therefore provide significant new constraints on the form of the non-perturbative contribution to the $p_T^W$ distribution.

Moreover, recent theoretical developments in combining the advantages of $b-$space and $p_T-$space resummation formalisms may provide a better theoretical framework for extracting the underlying $p_T^W$ distribution in Run II [ 17].

In short, the precision $Z$ data available in Run II together with further theoretical advances will reduce in a number of ways the systematic uncertainties due to the knowledge of the $p_T^W$ distribution, perhaps down to the level of $\sim 5$ MeV/c$^2$.

### 5.3. QCD Higher Order Effects

The $W$ bosons are treated as spin-one particles and decay via the weak interaction into a charged lepton ($e$, $\mu$ or $\tau$) and a neutrino. The charged leptons are produced from the $W$ decay with an angular distribution determined by the $\mathcal{O}(\alpha_s^2)$ calculation of [ 18] which, for $W^+$ bosons with a helicity of –1 with respect to the proton direction, has the form :

$$\frac{d\sigma}{d\cos\theta_{\rm CS}} \propto 1 + a_1(p_T)\cos\theta_{\rm CS} + a_2(p_T)\cos^2\theta_{\rm CS} \quad (7)$$

where $p_T$ is the transverse momentum of the $W$ and $\theta_{\rm CS}$ is the polar direction of the charged lepton with respect to the proton direction in the Collins-Soper frame [ 19]. $a_1$ and $a_2$ are $p_T$ dependent parameters. For $p_T = 0$, $a_1 = 2$ and $a_2 = 1$, providing the angular distribution of a $W$ boson fully polarized along the proton direction. For the $p_T^W$ values relevant to the $W$ mass analysis ($p_T^W < 30$ GeV/c), the change in $W$ polarization as $p_T^W$ increases only causes a modest change in the angular distribution of the decay leptons [ 18].

While the uncertainty associated with the change in the angular distribution of the $W$ decay lepton due to higher order QCD corrections (a few MeV/c$^2$) has been negligible for the Run I measurements, it can not be ignored for the Run II measurements (see the Photon and Weak Boson Physics working group report for more details).

### 5.4. QED Radiative Effects
### 5.4.1. Introduction

The understanding of QED radiative corrections is crucial for precision $W$ mass measurements at the Tevatron. The dominant process is final state radiation (FSR) from the charged lepton, the effect of which strongly depends on the lepton identification criteria and the energy or momentum measurement methods employed. Calorimetric energy measurements, such as those employed in the electron channel, are more inclusive than track based momentum measurements used in the muon channel and the effect of FSR is consequently reduced. In the CDF Run Ib $W$ mass measurement the mass shifts due to radiative effects were estimated to be $-65 \pm 20$ MeV/c$^2$ and $-168 \pm 10$ MeV/c$^2$ for the electron and muon channels, respectively [ 14]. These effects will be larger in Run II due to increase in tracker material in CDF and magnetic tracking in DØ.

The Monte Carlo program used for the Run I $W$ mass measurement incorporated a calculation of QED corrections by Berends and Kleiss [ 20]. This treatment, however, does not include initial state radiation (ISR) and has a maximum of one final state photon. The effect of multiple photon emission was estimated by comparing the calculation of Berends and Kleiss to PHOTOS [ 21], a universal Monte Carlo program for QED radiative corrections that can generate a maximum of two final state photons. Likewise, the effect of ISR and other missing diagrams was estimated by comparing the calculation of Berends and Kleiss to a full $\mathcal{O}(\alpha)$ calculation by Baur *et al.* [ 22]. The resulting systematic uncertainties on the $W$ mass are estimated to be 20 MeV/c$^2$ and 10 MeV/c$^2$ in the electron and muon channels, respectively [ 14]. Clearly these systematic uncertainties become much more significant in the context of statistical uncertainties of $\mathcal{O}(10)$ MeV/c$^2$ expected for 2 fb$^{-1}$ in Run II.

The next section describes in more detail the calculation by Baur *et al.*, which forms the basis for a new event generator. Some studies of the effects of QED radiation on the W mass measurement are presented in section 5.4.4. Section 5.4.5 briefly outlines some work in progress that should further reduce systematic uncertainties due to radiative corrections in Run II.

### 5.4.2. WGRAD

WGRAD is a program for calculating $\mathcal{O}(\alpha)$ electroweak radiative corrections to the process $q\bar{q}' \to W^{\pm} \to \ell^{\pm}\nu$, including the real photon contribution $q\bar{q}' \to \ell^{\pm}\nu\gamma$. Both ISR from the incoming quarks, FSR from the final state charged lepton, and interference terms are included. Many more details can be found in [ 22].

The most important generator level cuts are on the final state photon energy and collinearity for radiative events. The photon energy cut, controlled by the parameter $\delta_s$, is made on the fraction of the parton's energy carried by the emitted photon in the parton-parton center of mass system: $E_{\gamma}^* > \delta_s\sqrt{\hat{s}}/2$. The photon collinearity cut, controlled by the parameter $\delta_c$, is made on the angle between the charged fermion and the emitted photon in the same frame: $\cos\theta^* < 1 - \delta_c$. However, final state collinear singularities are regulated by the finite lepton masses and the above cut is only implemented for quarkonic radiation when ISR is included. The fraction of radiative events corresponding to different photon cuts is given in Table 4 for the process $q\bar{q}' \to \mu\nu(\gamma)$ at $\sqrt{s} = 2$ TeV. Loose fiducial cuts $p_T^\mu > 10$ GeV/c, $|\eta^\mu| < 2$ and $p_T^\nu > 10$ GeV/c have been applied. The inclusion of ISR increases the photon yield by around 30%, depending on the soft and collinear photon cuts applied. The fractions are significantly higher for the process $q\bar{q}' \to e\nu(\gamma)$ in the cases that FSR is included. The effect on the fitted $W$ mass of the inclusion of ISR is examined in section 5.4.4.

Table 4
The fraction of $q\bar{q}' \to \mu\nu(\gamma)$ events containing a final state photon for different final state photon soft and collinear cuts. Events are generated with ISR only, FSR only, and with a full treatment of QED radiation.

| Photon Cuts | ISR | FSR | Full |
|---|---|---|---|
| Photon Cuts | ISR | FSR | Full |
| $\delta_s = 0.01, \delta_c = 0.01$ | 1.6% | 9.4% | 11.1% |
| $\delta_s = 0.01, \delta_c = 0.001$ | 2.5% | 9.4% | 12.0% |
| $\delta_s = 0.001, \delta_c = 0.001$ | 4.1% | 15.5% | 20.0% |
| $\delta_s = 0.001, \delta_c = 0.0001$ | 5.2% | 15.5% | 21.3% |

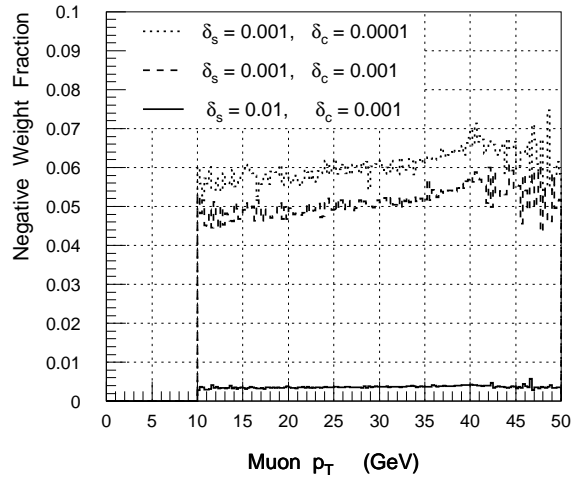

Figure 10. The negative weight fraction versus $p_T^\mu$ for different soft and collinear photon cuts.

### 5.4.3. Event Generation

WGRAD has been turned into an event generator through a suitable unweighting scheme described extensively in [ 23]. A significant complication is the presence of negative $q\bar{q}' \to \ell^\pm \nu$ event weights in the program which, while expected to cancel with positive $q\bar{q}' \to \ell^\pm \nu \gamma$ event weights in the calculation of physical observables, nevertheless appear separately in the unweighting procedure. The approach here has been to unweight the negative weight events in a similar manner to the positive weight events, such that the output consists of both positive and negative unit weight events. The fraction of negative weight events, plotted in Fig. 10 for the process $q\bar{q}' \to \mu\nu(\gamma)$, depends strongly on the soft and collinear photon cuts applied. It is not significantly different for $q\bar{q}' \to e\nu(\gamma)$ events. The effect of negative weights on the fitted $W$ mass is examined in the next section.

### 5.4.4. The Effect of QED Radiation on the Measured W Mass

WGRAD has been used to generate large $W \to \mu\nu$ event samples for the purposes of investigating the effect of QED radiation on the measurement of the $W$ mass. The events have been generated at $\sqrt{s} = 1.8$ TeV in order to make use of the CDF Run I $W$ production model and detector smearing parameterizations. The $W$ production model, extracted from the Run I Drell-Yan data, is used to smear the true $W$ transverse momentum. The CDF recoil model is then used to translate this into a measured $p_T(W)$, which is combined with the smeared lepton and photon momenta to form a realistic transverse mass distribution. Loose fiducial cuts $p_T^\mu > 20$ GeV/c, $|\eta^\mu| < 2$ and missing-$E_T > 20$ GeV are applied. To simulate the CDF muon identification criteria, events are rejected if a photon with $E_\gamma > 2.0$ GeV is found within an $\eta - \phi$ cone of radius 0.25 around the muon. Low energy photons inside the cone are not included in the measurement of the muon $p_T$, as is the case experimentally.

The unweighted event samples, all generated with $M_W = 80.4$ GeV/c$^2$ and $\Gamma_W = 2.1$ GeV, are divided into "data" and "Monte Carlo" sub-samples and fitted against one another in pseudo-experiments. The fit is to the transverse mass distribution in the range $50 < M_T < 100$ GeV/c$^2$. For a number of events in the transverse mass fit region equal to that in the CDF Run Ib $W \to \mu\nu$ analysis, the resulting statistical error is very similar.

As a cross check of this procedure, "data" and "Monte Carlo" samples generated with identical cuts are fitted against one another, with the result shown in Table 5(a). It is interesting to note that if the negative weight events, which occur at the 0.2% level in the "Monte Carlo" sample, are removed, the fit result changes by less than 0.5 MeV/c$^2$ (Table 5(b)).

Table 5
The results of pseudo-experiments generated by WGRAD with different treatments of QED radiative effects. Further details are given in the text.

|     | "Data" | "Monte Carlo" | Fit Result |
| --- | --- | --- | --- |
| (a) | FULL; $\delta_s = 0.01, \delta_c = 0.01$ | FULL; $\delta_s = 0.01, \delta_c = 0.01$ | $80.4030 \pm 0.0069$ GeV/c$^2$ |
| (b) | FULL; $\delta_s = 0.01, \delta_c = 0.01$ | FULL no neg.; $\delta_s = 0.01, \delta_c = 0.01$ | $80.4027 \pm 0.0069$ GeV/c$^2$ |
| (c) | FSR only; $\delta_s = 0.01, \delta_c = 0.01$ | FULL; $\delta_s = 0.01, \delta_c = 0.01$ | $80.392 \pm 0.006$ GeV/c$^2$ |
| (d) | FSR only; $\delta_s = 0.001, \delta_c = 0.001$ | FULL; $\delta_s = 0.01, \delta_c = 0.01$ | $80.381 \pm 0.006$ GeV/c$^2$ |
| (e) | FULL; $\delta_s = 0.001, \delta_c = 0.001$ | FULL; $\delta_s = 0.01, \delta_c = 0.01$ | $80.389 \pm 0.009$ GeV/c$^2$ |
| (f) | FULL no neg.; $\delta_s = 0.001, \delta_c = 0.001$ | FULL; $\delta_s = 0.01, \delta_c = 0.01$ | $80.399 \pm 0.009$ GeV/c$^2$ |

Table 5(c) shows the result of fitting "data" generated with FSR only. The shift in the fitted $W$ mass of $\approx 8$ MeV/c$^2$ is consistent with the estimate given in [14] of the effect of ISR on the fitted $W$ mass, although the uncertainties here are rather large. If the soft and collinear cuts are reduced in the "data" sample, as shown in Table 5(d), the fitted $W$ mass shifts significantly downwards. This is to be expected since the track based muon $p_T$ measurement does not incorporate collinear photons. The setting of soft and collinear photon cuts is therefore particularly important in the generation of $W \to \mu\nu$ Monte Carlo samples.

The fits shown in Table 5(e) and (f) are performed in order to examine the effect of negative weights on the fit when, as in the case of this "data" sample, negative weights are present at the 5% level. When the negative weight events are excluded from the fit, the result changes by 10 MeV/c$^2$. The larger shift in the fitted $W$ mass with respect to Table 5(b) is commensurate with the larger negative weight fraction in this sample.

### 5.4.5. Work in Progress

A remaining source of systematic uncertainty due to QED radiation is the effect of multiple photon emission. As discussed above, this has previously been estimated by comparing the Berends and Kleiss single photon calculation with the results of running the PHOTOS algorithm. Recently, however, complete matrix element calculations of the processes $q\bar{q}' \to \ell^\pm \nu \gamma\gamma$ and $q\bar{q} \to \ell^+\ell^- \gamma\gamma$ have been performed [24]. It may be possible in the future to do detailed comparisons of the results of these calculations and the PHOTOS algorithm, in order to arrive at a better constrained systematic uncertainty due to multiple photon emission.

Furthermore, a complete set of $\mathcal{O}(\alpha)$ electroweak radiative corrections to the process $q\bar{q} \to \ell^+\ell^-$, including the real photon contribution $q\bar{q} \to \ell^+\ell^-\gamma$, will soon be available. This will enable a consistent Monte Carlo description of the $W$ data and the $Z$ data, upon which the $W$ mass analysis crucially depends for the understanding of gauge boson production and the calibration of the detectors.

### 5.4.6. Summary and Conclusions

Systematic uncertainties due to QED radiative effects currently run at the level of $\approx 20$ MeV/c$^2$ in the electron channel and $\approx 10$ MeV/c$^2$ in the muon channel. A large contribution to this uncertainty is the effect of ISR and interference terms, which are not present in the Berends and Kleiss calculation and the PHOTOS algorithm that have previously been used in $W$ production Monte Carlo programs.

A full $\mathcal{O}(\alpha)$ calculation by Baur *et al.* has been used as the basis for a new event generator. The results of several pseudo-experiments generated with different treatments of QED radiative effects agree with previous estimates. They show that negative weights need to be treated carefully, especially in the case of very small soft and collinear photon cuts.

Further studies of QED radiative corrections to $W$ production will continue as new calculations become available. It is clear, however, that the use of new programs such as WGRAD could significantly reduce systematic uncertainties due to QED radiative corrections in Run II, either through explicit corrections being applied to the extracted $W$ mass, or through their use in new Monte Carlo event generators. The remaining systematic uncertainties due to QED corrections might then be reduced to the level of 5 MeV/$c^2$ and 10 MeV/$c^2$ in the muon and electron channels, respectively.

## 6. Summary of Run II Expectations

As has been discussed in previous sections, many of the systematic uncertainties in the $W$ mass measurement approximately scale with statistics. These are listed in Table 6 for the Run Ib CDF muon analysis and should scale to $\approx$ 20 MeV/$c^2$ for an integrated luminosity of 2 fb$^{-1}$. With reasonable assumptions for the size of non-scaling systematics such as those due to PDFs and higher order QED effects, a 40 MeV/$c^2$ measurement in the muon channel by each experiment seems achievable. The systematic uncertainties in the electron channel are less easy to extrapolate given the particular sensitivity to calorimeter scale non-uniformities in this channel and the extra material in the Run II tracking detectors. The detailed understanding of detector performance is of course difficult to anticipate, although it is clear that both scalable and non-scaling systematics would be easier to understand if fast Monte Carlo generators including all the relevant effects were available.

The individual uncertainties for the Run Ib $\Gamma_W$ measurement are listed in Table 7 together with their projections for 2 fb$^{-1}$. All but the last three sources of error are constrained directly from collider data, and hence should scale roughly as $1/\sqrt{\mathcal{L}}$. While the last three uncertainties may decrease somewhat as new measurements and calculations become available, they will not scale statistically with the Run II dataset. Assuming no improvement in these three uncertainties, while all others scale statistically, each experiment can make a $\sim 40$ MeV width measurement, combining $e$ and $\mu$ channels for a 2 fb$^{-1}$ dataset.

Table 6
Errors on the CDF Run Ib muon $W$ mass which scale statistically, in MeV/$c^2$.

| Source | error |
| --- | --- |
| Fit statistics | 100 |
| Recoil model | 35 |
| Momentum resolution | 20 |
| Selection bias | 18 |
| Background | 25 |
| Momentum scale | 85 |

Table 8
Dominant uncertainties for contrasting components of the DØ $M_W$ determination. The quantities shown are the shift in $M_W$ for a $1\sigma$ change in the relevent parameter. The EM resolution term refers to the sampling term for the resolution function. Taken from Ref. [25].

| Source | $\delta M_W(M_T)$ | $\delta M_W(p_T^e)$ |
| --- | --- | --- |
| $p_T^W$ | 10 | 50 |
| EM resolution | 23 | 14 |
| hadron scale | 20 | 16 |
| hadron resolution | 25 | 10 |
| backgrounds | 10 | 20 |

## 7. Other Methods of Determining $M_W$ at the Tevatron

While the traditional transverse mass determination has been the optimal technique for the extraction of $M_W$ in the low-luminosity running at hadron colliders, other techniques have been or may be employed in the future. These methods may shuffle or cancel some of the systematic and statistical uncertainties resulting in more precise measurements.

### 7.1. Transverse Momentum Fitting

As noted above, the most obvious extensions of the traditional transverse mass approach to determining $M_W$ are fits of the Jacobian kinematical edge from the transverse momentum of both leptons. DØ has measured $M_W$ using all three distributions and the uncertainties are indeed ordered as one would expect: The fractional uncertainties on $M_W$ from the DØ Run I measurements for the three methods of fitting are: 0.12% ($M_T$), 0.15% ($p_T^e$), and 0.21% ($p_T^\nu$). As expected, the $p_T^e$ method is slightly less precise than the transverse mass. However, for a central electron ($|\eta| < 1$), the uncertainty in the $p_T^e$ measurement due to the $p_T^W$ model is 5 times that in the $M_T$ measurement. As can be seen from Table 8, this is nearly balanced by effects from electron and hadron response and resolutions which are relatively worse for $M_T$. Accordingly, when there are sufficient statistics to enable cuts on the measured hadronic recoil, the measurement uncertainty from the $p_T^W$ model might be better controlled and enable the $p_T^e$ measurement to compete favorably with the $M_T$ measurement which relies so heavily on modeling of the hadronic recoil. In order to optimize the advantages of all three measurements, the DØ final Run I determination of $M_W$

Table 7
Sources of error for CDF 1994-95 $\Gamma_W$ measurement and extrapolations to 2 fb$^{-1}$. The last three uncertainties are common to the $e$ and $\mu$ analyses.

|  | CDF 1994-95 ($\to$ 2fb$^{-1}$) | |
| --- | --- | --- |
| Source | $\Delta\Gamma$ ($e$, MeV) | $\Delta\Gamma$ ($\mu$, MeV) |
| Statistics | 125 ($\to$ 30) | 195 ($\to$ 45) |
| Lepton $E$ or $p_T$ non-linearity | 60 | 5 |
| Recoil model | 60 | 90 |
| $W$ $P_T$ | 55 | 70 |
| Backgrounds | 30 | 50 |
| Detector modeling, lepton ID | 30 | 40 |
| Lepton $E$ or $p_T$ scale | 20 | 15 |
| Lepton resolution | 10 | 20 |
| PDFs (common) | 15 | 15 |
| $M_W$ (common) | 10 | 10 |
| QED (common) | 10 | 10 |
| Uncorrelated systematic | 112 ($\to$ 25) | 133 ($\to$ 30) |
| Correlated systematic | 21 | 21 |
| Total systematic | 115 ($\to$ 33) | 135 ($\to$ 37) |
| Total stat + syst | 170 ($\to$ 45) | 235 ($\to$ 60) |

combined the separate results [25].

The resolution sensitivity for muon measurements is even less than that for electrons so that has the benefit of slightly favoring a transverse mass measurement with muons over that for electrons.

### 7.2. Ratio Method

DØ has preliminarily determined $M_W$ by consideration of ratios of $W$ and $Z$ boson distributions which are correlated with $M_W$ [26]. The principle is that one can cancel common scale factors in ratios and directly determine the quantity $r^{meas} \equiv \frac{M_W}{M_Z}$, which can be compared with the precise LEP $M_Z^{LEP}$. The quantities that have been considered are:

1. $r(M_T)$ and $r(p_T)$, which has the advantage of being well-studied [27]. There are challenges with this approach which will be discussed below.

2. $r(E^e)$ which has the advantage that the peak of the distribution is precisely correlated with $M_W$, but the disadvantage that statistical uncertainty washes out the position of that peak.

3. The difference of transverse mass distributions (not as precise as ratios).

The procedure is to compare two distributions, one for $W$ bosons and a similarly constructed one for $Z$ bosons, for example, $f^{W,Z}(x)$ as a function of a given variable, such as $x = M_T$ or $x = p_T^e$. Practically speaking, the $Z$ boson decay electrons are scaled by a factor $s$ and $f^Z(x, s)$ is compared with $f^W(x)$ as a function of $x$, for different trial values of $s$. A statistical measure (the Kolmogorov-Smirnov test) is calculated for each $s$ and the value of the highest Kolmogorov-Smirnov probability, $s^{best}$, is declared to be $r^{meas}$ and the desired mass is then extracted from $M_W = r^{meas} \times M_Z^{LEP}$. In principle, minimal Monte Carlo fitting is required, as the measurement is performed with data.

Figure 11 shows the idea with an unsmeared $Z$ boson transverse mass distribution compared to a simulated (unsmeared) $W$ boson distribution. Various values of $s$ lead to various mismatches between $f^W(M_T)$ and $f^Z(M_T)$ which can be characterized by a Kolmogorov-Smirnov probability as a function of $M_W = s \times M_W^{LEP}$. This probability distribution for an ensemble of 100 Monte Carlo experiments is shown in Fig. 12 resulting in an RMS of 40 MeV/c$^2$.

However, there are challenges to be faced using this technique.

- Many systematic effects cancel in this method, such as electromagnetic scale, hadronic scale, angular scale, luminosity effects. However, these are first-order cancellations, some of which in the end are not sufficient: the second order effects from these quantities must be considered. Likewise, most resolutions have additive terms which do not cancel in a ratio.

- The statistical precision of the $Z$ sample is directly propagated into the resultant overall $\delta M_W$, in contrast to the traditional approach where the $Z$ boson statistics is a component of various of the measured resolutions.

- The detector modeling must take into account

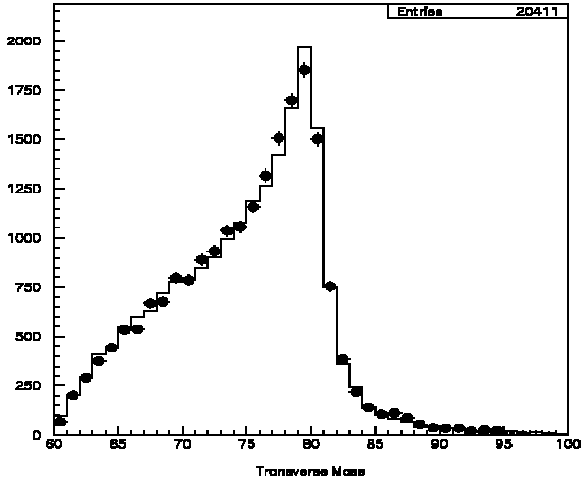

Figure 11. For unsmeared Monte Carlo events, the transverse mass of simulated $W$ bosons (histogram) is overlayed with that of $Z$ boson (dots) events in which the electron has been scaled by a factor which produces the best match.

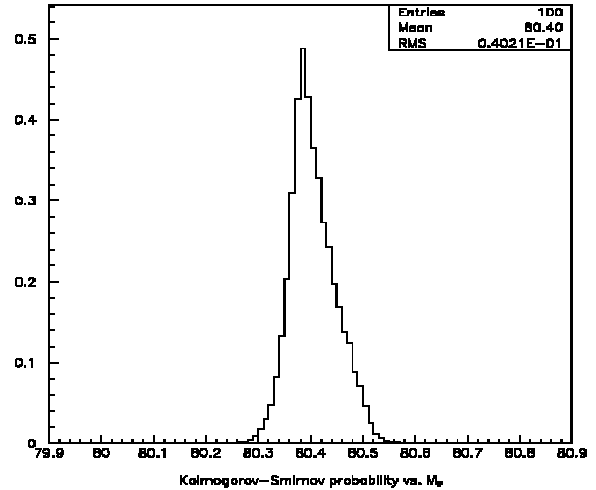

Figure 12. The Kolmogorov-Smirnov probability distribution for various scales in a comparison of $W$ and $Z$ boson unsmeared events corresponding to 100 experiments. The RMS is 40 MeV/c$^2$ for 20000 events.

small, but important differences between $Z$ and $W$ events such as underlying event, resolutions, efficiencies, acceptances, and the effects of the "extra" electron in $Z$ boson events which complicates underlying event and recoil measurements.

- From the physics model, there are also differences between the two samples which must be considered, such as the fact that the production of $Z$ and $W$ bosons take place from the annihilation of like and unlike flavored quarks, respectively and that weak asymmetries lead to different decay angular distributions.

- Particularly difficult is the need to "extra-smear" the electrons from $Z$ boson decays. This is due to the fact that $p_T^e$ values for the heavier $Z$ boson are harder, resulting in a different average resolution smearing. This same effect is true for the recoil distributions between $Z$ and $W$ bosons.

- Finally, the acceptances for the two bosons are different since there are potentially two opportunities to select a $Z$ boson event at the trigger and event selection stages. Similarly, there is an acceptance difference in the opposite direction due to electrons in $Z$ bosons being lost in cracks between the CC and EC calorimeters in the DØ detector.

An analysis from Run Ia data from the DØ experiment has been done [26]. Figure 13 shows data for the scaled comparison and the unscaled original distributions. Electrons from the $W$ boson events were selected to have $p_T^e > 30$ GeV/c, while those from $Z$ boson events, must satisfy $p_T^e > 34.1$ GeV/c. Electrons from the $W$ sample and at least one electron from the $Z$ samples were required to be in the central calorimeter. This results in 5244 $W$ bosons and 535 $Z$ boson events. Backgrounds are subtracted according to the traditional analysis. "Extra-smearing" is done for each accepted $Z$ boson event (twice, for both electrons) 1000 times, using a different random seed for each smearing. Differences in the $W$ and $Z$ boson production mechanisms and acceptances result in an effective correction of 109 MeV/c$^2$, while the difference in radiative corrections results in an effective correction of $-116$ MeV/c$^2$. The magnitude of these corrections is not very different from corrections within the traditional technique and the demand on knowing the uncertainties in them is similarly stringent. Figure 14 shows the probability distribution for the result. The preliminary result from this analysis for central, Run Ia electrons is

$$M_W = 80.160 \pm 0.360 \pm 0.075 \text{ GeV/c}^2.$$

Comparison with the traditional Run Ia result from the same data is readily made, but most appreciated

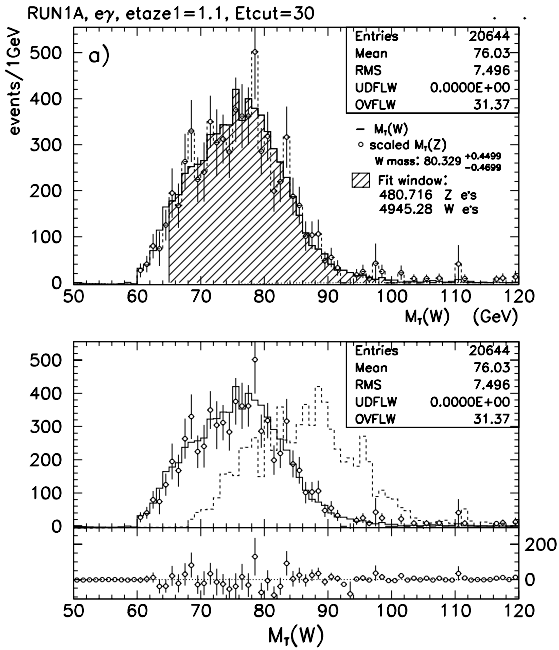

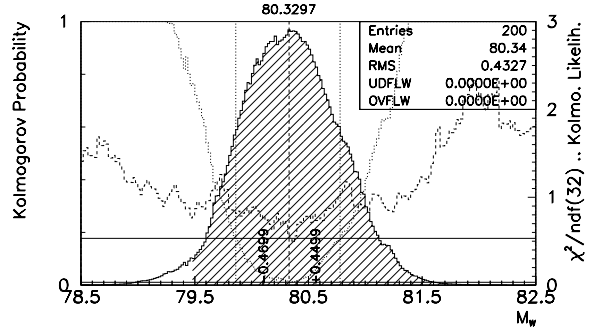

Figure 14. The Kolmogorov probability distribution (hatched) is shown as a function of the $W$ boson mass used as a scale factor. The dotted curve is the Kolmogorov Likelihood and the dashed curve is the $\chi^2/\mathrm{ndf}$ distribution (right axis).

Figure 13. (a) The transverse mass distribution of $W$ (solid) and scaled $Z$ (dots) bosons is shown along with the hatched fit window. (b) The original distributions are shown, along with their difference. The $Z$ distribution has been normalized to that of the $W$ boson sample.

with a slightly different interpretation of the Run Ia uncertainties. The Run Ia result [28] from Table 2 is

$$80.350 \pm 0.140 \pm 0.165 \pm 0.160 \text{ GeV}/c^2$$

where the first error is the statistical uncertainty (from $W$ events), the second is the systematic uncertainty and the third is the electron scale determination. It is important to note that the scale uncertainty is almost completely dominated by the $Z$ boson statistics. Therefore, as a statistical uncertainty, it can be combined with the $W$ uncertainty of 140 MeV/$c^2$ for the purposes of comparison with the ratio method. This results in an overall "statistical" uncertainty of 212 MeV/$c^2$. Now, the stronger systematic power of the ratio method is apparent (75 versus 165 MeV/$c^2$) and the poorer statistical power (360 versus 212 MeV/$c^2$) is also evident.

### 7.2.1. Prospects for Run II

This apparent systematic power of the ratio method can only fully be realized in high luminosity running, such as Run II. The ratio method analysis of the DØ Run Ib data was recently completed [29]. The Run Ib sample has 82 pb$^{-1}$ of data (1994–1995 data set), 33,137 $W$ and 4,588 $Z$ events (electrons in both Central and End Calorimeters of DØ) after the standard electron selection cuts. The $W$ mass resulting from the ratio fit is $M_W = 80.115 \pm 0.211$ (stat.) $\pm$ 0.050 (syst.) GeV/$c^2$. The statistical uncertainty is in good agreement with an ensemble study of 50 Monte Carlo samples of the same size ($80.36 \pm 0.25$ GeV/$c^2$).

Early efforts at predicting the results for a Run II sample of 100,000 $W$ bosons is shown in Fig. 15 with full detector acceptances and resolutions taken into account. The statistical precision from this fit is of the order of 20 MeV/$c^2$ and the systematic uncertainties may be nearly negligible.

## 8. Prospects for Measuring $M_W$ at Other Accelerators

### 8.1. LEP II

The prospects for determination of $M_W$ at LEP II have become fully understood in the last year with the accumulation of hundreds of pb$^{-1}$ at four center of mass energies. Here we review the status as of the Winter 2000 conferences and project the prospects through to the completion of electron-positron running at CERN. For a review, see Ref. [30, 31].

#### 8.1.1. Data Accumulation

The annihilation of $e^+e^-$ into $W$ boson pairs occurs via three diagrams: a $t$-channel neutrino exchange and $s$-channel $Z$ or $\gamma$ exchange. The final states from the decays of the two $W$ bosons are: both $W$ bosons decay into hadrons ($qqqq$, "4-q" mode); one $W$ decays into quarks, and the other into leptons ($e$, $\mu$, $\tau$ and their neutrinos, "$qq\ell\nu$"); and both $W$ bosons decay leptonically. Collectively, the latter two modes are referred to as "non-q" modes. The efficiencies and sample purities are typically quite high, as shown in Table 9.

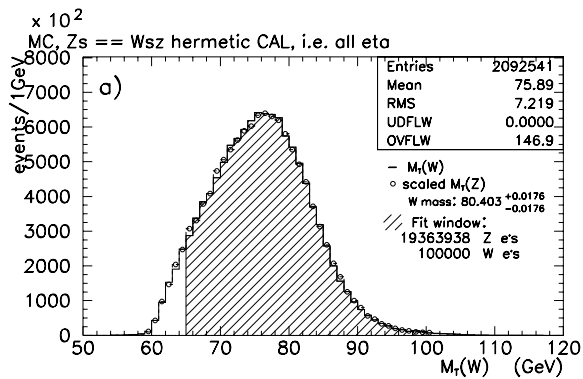

Figure 15. Same as Fig. 13a, but for 100,000 $W$ bosons.

Table 9
Efficiencies and sample purity for a representative LEP II experiment (OPAL).

| Channel | $\epsilon$ (%) | purity (%) |
|---|---|---|
| $qqqq$ | 85 | 80 |
| $qqe\nu$ | 85 | 80 |
| $qq\mu\nu$ | 87 | 80-90 |
| $qq\tau\nu$ | 66 | 80 |

The results by Spring 2000 come from running at center of mass energies: 172, 183, 189, and several energies between 190 GeV and 200 GeV. There is recent running above 200 GeV for a total of more than 400 pb$^{-1}$ accumulated per experiment. $M_W$ results are final for all four experiments for the 172 and 183 GeV sets [ 32, 33, 34, 35] and preliminary for the 1998 189 GeV running [ 36, 37, 38, 39]. In addition, ALEPH [ 40], L3 [ 41], and OPAL [ 42] have preliminary results from the collection of runs in the range from 190 GeV to 200 GeV. Table 10 shows the approximate accumulated running to date (July 2000).

There are broadly two methods employed for determining $M_W$ at LEP II. The first method is the measurement of the threshold of the $WW$ cross section and the second is the set of constrained fits possible for the various measured final states. The latter set of methods constitute the prominent results and employ construction of invariant masses making use of the beam constraints. There are a variety of methods, some of which make use of the constraint $M_{W_1} = M_{W_2}$ and some of which involve sophisticated multivariate analyses. The spirit of approach is much like the strategies employed in the top quark mass analyses of CDF and DØ.

Table 10
Approximate accumulated running per experiment. The 2000 totals are current as of the first week in July 2000.

| year | beam energy (GeV) | $\int \mathcal{L} dt$ (pb$^{-1}$) |
|---|---|---|
| 1996 | 80.5-86 | 25 |
| 1997 | 91-92 | 75 |
| 1998 | 94.5 | 200 |
| 1999 | 96-102 | 250 |
| 2000 | 100-104 | 100 |

Table 11
Typical systematic uncertainties on $M_W$ for a generic LEP II experiment from data sets corresponding to the 189 GeV running. Entries are approximate and broad averages meant to give a relative sense of scale only. The Source labels are generally self-explanatory, with CR/BE standing for "Color Reconnection" and "Bose-Einstein" respectively.

| Source | $qq\ell\nu$ (MeV/c$^2$) | $qqqq$ (MeV/c$^2$) |
|---|---|---|
| ISR | 15 | 15 |
| frag | 25 | 30 |
| 4 fermion | 20 | 20 |
| detector | 30 | 35 |
| fit | 30 | 30 |
| bias | 25 | 25 |
| bckgrnd | < 5 | 15 |
| MC stat | 10 | 10 |
| Subtotal | 61 | 67 |
| LEP | 17 | 17 |
| CR/BE | | 60 |
| Total | 63 | 85 |

The results are treated separately for the $qq\ell\nu$ and $qqqq$ final states due to the significant differences in systematic uncertainties. Typical uncertainty contributions are listed in Table 11 [ 43]. Many of the experimental uncertainties, such as scale, background, and Monte Carlo generation, are statistically limited. For example, there is a fixed amount of $\sqrt{s} = M_Z$ running in each running period and that contributes a statistical component to the energy scale uncertainty.

The dominant uncertainty comes from the final state effects in the *qqqq* channel. Because the outgoing quarks can have color connections among them, the fragmentation of the ensemble of quarks into hadrons are not independent. This leads to an theoretical uncertainty called "Color Reconnection" (CR). In addition, since the hadronization regions of the $W^+$ and $W^-$ overlap, coherence effects between identical low-momentum bosons originating from different $W$'s due to Bose-Einstein (BE) correlations may be present. The combined total of these two effects is currently accepted to contribute 52 MeV/$c^2$ of uncertainty to the *qqqq* results. Ultimately, the non-CR/BE uncertainty will likely be the uncertainty in modeling single-quark fragmentation and associated QCD emission effects.

### 8.1.2. Results, April 2000

The preliminary results for $M_W$ from the combined data taking through 1999 running period are shown in Table 12. The combined LEP result for the $qq\ell\nu$ channels is [44]:

$$M_W^{qq\ell\nu} = 80.398 \pm 0.039 \pm 0.031 \pm 0.017 \text{ GeV}/c^2$$

where the first error is statistical, the second systematic, and the third the LEP energy scale. The combined preliminary result for the *qqqq* channel is:

$$M_W^{4q} = 80.408 \pm 0.037 \pm 0.031 \pm 0.016 \pm 0.052 \text{ GeV}/c^2$$

where the first three errors are the same as for the $qq\ell\nu$ result and the fourth error is due to the combined CR/BE theoretical uncertainty. Taking into account the correlations, the combined preliminary result from constrained fitting for all channels is:

$$M_W^{4f} = 80.401 \pm 0.027 \pm 0.031 \pm 0.017 \pm 0.018 \text{ GeV}/c^2$$

where the four errors are in the same order as for the *qqqq* result. The current overall result comes from combining the above with that from the threshold measurement of

$$M_W^{\sigma(E)} = 80.400 \pm 0.220 \pm 0.025 \text{ GeV}/c^2.$$

Here the first error is combined statistical and systematic and the second error is the error due to LEP energy scale. This results in the preliminary overall LEP II (April 2000) value of

$$M_W^{LEP} = 80.401 \pm 0.048 \text{ GeV}/c^2.$$

### 8.1.3. Prospects for the Future

The current results are preliminary and running is underway at this writing with the end of LEP II scheduled for the beginning of October, 2000. Eventually, the 1999 data will be fully analyzed and, with the accumulation of the final 2000 running, should result in a combined statistical and systematic uncertainty (excluding the CR/BE and LEP contributions) of approximately 35 MeV/$c^2$ [1]. With the overall contribution of 18 MeV/$c^2$ and 17 MeV/$c^2$ from the CR/BE and LEP errors respectively, the ultimate limit from LEP II $W$ boson pair determination of $M_W$ should be approximately 40 MeV/$c^2$.

Table 12
Preliminary LEP II results for $M_W$ by experiment and according to reconstructed channel. The results are from the combination of 1996-1998 running (all experiments) plus preliminary results from 1999 running for ALEPH, DELPHI, L3, and OPAL.

| Experiment | $M_W$ (GeV/$c^2$) | |
|---|---|---|
| | $qq\ell\nu$ | $qqqq$ |
| ALEPH | $80.435 \pm 0.079$ | $80.467 \pm 0.086$ |
| DELPHI | $80.230 \pm 0.140$ | $80.360 \pm 0.115$ |
| L3 | $80.282 \pm 0.102$ | $80.489 \pm 0.132$ |
| OPAL | $80.483 \pm 0.078$ | $80.380 \pm 0.103$ |

### 8.2. LHC

It was pointed out several years ago [45] that the LHC has the potential to provide an even more precise measurement of $M_W$. This suggestion was based on the observations that the precision measurement of $M_W$ at hadron colliders has been demonstrated to be possible; that the statistical power of the LHC dataset will be huge; and that triggering will not be a problem. These authors estimated that $M_W$ could be determined to better than 15 MeV/$c^2$. More recently, the ATLAS collaboration has studied the question in more detail [46] and arrived at an uncertainty of 25 MeV/$c^2$, per experiment, in the electron channel alone.

Achieving such precision will require substantial further reduction of theoretical and systematic uncertainties, all of which must be reduced to the $\leq$ 10 MeV level. This includes the contributions from the $W$ production model, parton distributions, and radiative decays, as well as experimental systematics such as the energy-momentum scale of the detector and any complications from underlying energy deposition even in the low luminosity running. While some have questioned whether such "heroic" progress will ever be possible, we would argue that it is futile to debate the question at this time. Rather, the best indicator of future LHC precision will be to see how well the Fermilab experiments manage to deal with the significant improvements in systematics which will be necessary in order to match the anticipated precision for $m_t$. The point to be made is this: should it prove necessary to determine the $W$ mass to a precision of 10–20 MeV/$c^2$, the LHC will have the *statistical power* to continue the hadron collider measurements into this domain. The success of such a program will then depend on

- Consensus in the field that such precision is needed. One such justification might be to distinguish among different models of supersymmetry-breaking using global fits including $M_W$, the top mass and the light Higgs mass. It is likely that a big parallel effort to push down the top mass uncertainty to the 1 GeV/c$^2$ level would also then be needed;

- A major, multi-year effort within the LHC experiments in order to understand their detectors and their response to leptons, missing transverse energy and recoil hadrons at the required level. This is a measurement which places heavy burdens on manpower as it requires an understanding of the detector which is more precise than for any other measurement;

- A comparable major effort to reduce the theoretical uncertainties through better calculations, through control-sample measurements, and work on parton distributions.

This is not a program that will be undertaken lightly. But should it turn out to be necessary, the experience of Run II at the Tevatron will be invaluable in carrying it out.

### 8.3. A Linear Collider

The $W$ mass can be measured at a Linear Collider (LC) in $W^+W^-$ production either in a dedicated threshold scan operating the machine at $\sqrt{s} \approx 161$ GeV, or via direct reconstruction of the $W$ bosons in the continuum ($\sqrt{s} = 0.5 - 1.5$ TeV). Both strategies have been used with success at LEP II.

In the threshold region, the $W^+W^-$ cross section is very sensitive to the $W$ mass. The sensitivity is largest in the region around $\sqrt{s} = 161$ GeV [47] at which point the statistical uncertainty is given by

$$\delta M_W^{stat} \approx 90 \text{ MeV/c}^2 \left[ \frac{\varepsilon \int \mathcal{L} dt}{100 \text{ pb}^{-1}} \right]^{-1/2}. \quad (8)$$

Here, $\varepsilon$ is the efficiency for detecting $W$ bosons. For $\varepsilon = 0.67$ and an integrated luminosity of 100 fb$^{-1}$, one finds from Eq. (8)

$$\delta M_W^{stat} \approx 3.5 \text{ MeV/c}^2. \quad (9)$$

Assuming that the efficiency and the integrated luminosity can be determined with a precision of $\Delta\varepsilon = 0.25\%$ and $\Delta\mathcal{L} = 0.1\%$, $M_W$ can be measured with an uncertainty of [48]

$$\delta M_W \approx 6 \text{ MeV/c}^2, \quad (10)$$

**provided** that the theoretical uncertainty on the $W^+W^-$ cross section is smaller than about 0.1% in the region of interest.

Presently, the $W$ pair cross section in the threshold region is known with an accuracy of about 1.4% [49]. In order to reduce the theoretical uncertainty of the cross section to the desired level, the full $\mathcal{O}(\alpha)$ electroweak corrections in the threshold region are needed. This calculation is extremely difficult. In particular, currently no practicable solution of the gauge invariance problem associated with finite $W$ width effects in loop calculations exists. The existing calculations which take into account $\mathcal{O}(\alpha)$ electroweak corrections all ignore non-resonant diagrams [50].

If one (pessimistically) assumes that the theoretical uncertainty of the cross section will not improve, the uncertainty of the $W$ mass obtained from a threshold scan is completely dominated by the theoretical error, and the precision of the $W$ mass is limited to [47]

$$\delta M_W \approx \delta M_W^{theor} \approx 17 \text{ MeV/c}^2 \left[ \frac{\Delta\sigma}{\sigma} \times 100\% \right] \quad (11)$$
$$\approx 24 \text{ MeV/c}^2.$$

Using direct reconstruction of $W$ bosons and assuming an integrated luminosity of 500 fb$^{-1}$ at $\sqrt{s} = 500$ GeV, one expects a statistical error of $\delta M_W^{stat} \approx 3.5$ MeV/c$^2$ [51]. Systematic errors are dominated by jet resolution effects. Using $Z\gamma$, $Z \to 2$ jet events where the photon is lost in the beam pipe for calibration, a systematic error $\delta M_W^{syst} < 10$ MeV/c$^2$ is expected to be achieved. The resulting overall precision of the $W$ boson mass from direct $W$ reconstruction at a Linear Collider operating at an energy well above the $W$ pair threshold is

$$\delta M_W \approx 10 \text{ MeV/c}^2. \quad (12)$$

### 9. Theoretical Issues at high $\sqrt{s}$

Future hadron and lepton collider experiments are expected to measure the $W$ boson mass with a precision of $\delta M_W \approx 10 - 20$ MeV/c$^2$. For values of $\delta M_W$ smaller than about 40 MeV/c$^2$, the precise definition of the $W$ mass and width become important when these quantities are extracted.

In a field theoretical description, finite width effects are taken into account in a calculation by resumming the imaginary part of the $W$ vacuum polarization. This leads to an energy dependent width. However, the simple resumming procedure carries the risk of breaking gauge invariance. Gauge invariance works order by order in perturbation theory. By resumming the self energy corrections one only takes into account part of the higher order corrections. Apart from being theoretically unacceptable, breaking gauge invariance may result in large numerical errors in cross section calculations.

In order to restore gauge invariance, one can adopt the strategy of finding the minimal set of Feynman diagrams that is necessary for compensating those terms caused by an energy dependent width which violate gauge invariance [52]. This is relatively straightforward for a simple process such as $q\bar{q}' \to W \to \ell\nu$ [53], but more tricky for $e^+e^- \to W^+W^- \to 4$ fermions, in particular when higher order corrections are included. The so-called *complex mass* scheme [54], which uses a constant, *ie.* an energy independent width, offers a convenient alternative. At LEP II energies, $\sqrt{s} \approx 200$ GeV, the differences in the $e^+e^- \to 4$ fermions cross section using an energy dependent and a constant width are small. However, at Linear Collider energies, $\sqrt{s} = 0.5 - 2$ TeV, the terms associated with an energy dependent width which break gauge invariance lead to an overestimation of the cross section by up to a factor 3 [54].

For $q\bar{q}' \to W \to \ell\nu$, the parameterizations of the $W$ resonance in terms of an energy dependent and a constant $W$ width are equivalent. The $W$ resonance parameters in the constant width scenario, $\overline{M_W}$ and $\overline{\Gamma_W}$, and the corresponding quantities, $M_W$ and $\Gamma_W$, of the parameterization using an energy dependent width are related by a simple transformation [55]

$$\overline{M_W} = M_W \left(1 + \gamma^2\right)^{-1/2}, \quad (13)$$

$$\overline{\Gamma_W} = \Gamma_W \left(1 + \gamma^2\right)^{-1/2}, \quad (14)$$

where $\gamma = \Gamma_W/M_W$. The $W$ mass obtained in the constant width scenario thus is about 27 MeV/c$^2$ smaller than that extracted using an energy dependent width.

In the past, an energy dependent $W$ width has been used in measurements of the $W$ mass at the Tevatron [56, 57]. The Monte Carlo programs available for the $W$ mass analysis at LEP II (see Ref. [50] for an overview) in contrast use a constant $W$ width. Since the difference between the $W$ mass obtained using a constant and an energy dependent width is of the same size or larger than the expected experimental uncertainty, it will be important to correct for this difference in future measurements.

**10. Conclusions**

The measurements of the $W$ mass and width in Run I already represent great experimental achievements and contribute significantly to their world average determinations. Close inspection of the various systematic error sources leads us to believe that a $W$ mass measurement in Run II at the 30 MeV/c$^2$ level per experiment is achievable, and this compares well to the expected uncertainty on the $W$ mass measured at LEP II. Each experiment is expected to measure the $W$ width to a similar precision with 2 fb$^{-1}$ of data.

Alternative methods for determining $M_W$ at the Tevatron have been discussed and may turn out to be more appropriate in the Run II operating environment than the traditional transverse mass fitting approach. Determination of the $W$ mass at the LHC will be extremely challenging, using detectors that are not optimized for this measurement. A future linear collider should do significantly better. Clearly, the $W$ mass and width measurements at the Tevatron in Run II will remain the best hadron collider determinations of these quantities for many years and will compete with the best measurements made elsewhere.

# Measurement of the Forward-Backward Asymmetry in $e^+e^-$ and $\mu^+\mu^-$ events with DØ in Run II

Ulich Baur[a], John Ellison[b] and John Rha[b]

[a]State University of New York, Buffalo, NY 14260

[b]University of California, Riverside, CA 92521

The forward-backward asymmetry of $\ell^+\ell^-$ events in Run II can yield a measurement of the effective weak mixing angle $\sin^2\bar\theta_W$ and can provide a test of the standard model $\gamma^*/Z$ interference at $\ell^+\ell^-$ invariant masses well above the 200 GeV center of mass energy of the LEP II collider. The asymmetry at large partonic center of mass energies can also be used to study the properties of possible new neutral gauge bosons. We describe an updated study of the forward-backward asymmetry and give estimates of the statistical and systematic uncertainties expected in Run II. The prospects for measuring the weak mixing angle at the LHC and a linear collider operating at $\sqrt{s}=M_Z$ are also briefly described.

## 1. Introduction

In this note we present an updated study of the prospects for measurement of the forward-backward asymmetry in $p\bar p \to \gamma^*/Z \to \ell^+\ell^-$ events. This work extends our earlier study described in the TeV2000 report [1] in several respects: (i) we include the effects of QED corrections; (ii) we include the effects of expected Run II DØ detector resolutions and efficiencies; (iii) we consider systematic errors in more detail; and (iv) we include a simulation of the muon channel process $p\bar p \to \gamma^*/Z \to \mu^+\mu^-$.

The forward-backward asymmetry ($A_{FB}$) in $p\bar p \to \gamma^*/Z \to \ell^+\ell^-$ events arises from the parton level process $q\bar q \to \gamma^*/Z \to \ell^+\ell^-$. This asymmetry depends on the vector and axial-vector couplings of the quarks and leptons to the $Z$ boson and is therefore sensitive to the effective weak mixing angle $\sin^2\bar\theta_W$. The current world average value of $\sin^2\bar\theta_W$ from LEP and SLD asymmetry measurements is $\sin^2\bar\theta_W = 0.23147 \pm 0.00017$ [2]. As will be seen from our results it will be necessary to achieve high luminosity ($> 10$ fb$^{-1}$) and combine the results from the electron and muon channels and the results from DØ and CDF to achieve a precision comparable to this.

The SM tree level prediction [3] for $A_{FB}$ as a function of $\hat s$ for $q\bar q \to \gamma^*/Z \to e^+e^-$ is shown in Fig. 1 for $u$ and $d$ quarks. These are the same asymmetries as encountered in the inverse $e^+e^-$ annihilation reactions. The largest asymmetries occur at parton center-of-mass energies of around 70 GeV and above 110 GeV. At the $Z$-pole the asymmetry is dominated by the couplings of the $Z$ boson and arises from the interference of the vector and axial components of its coupling. The asymmetry is proportional to the deviation of $\sin^2\bar\theta_W$ from $\frac{1}{4}$. At large invariant mass, the asymmetry is dominated by $\gamma^*/Z$ interference and is almost constant ($\approx 0.6$), independent of invariant mass.

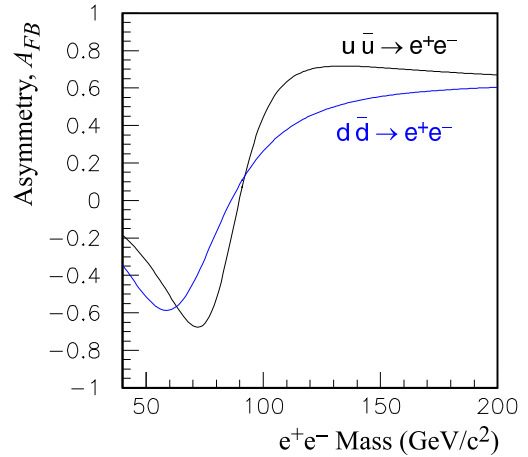

Figure 1. The standard model tree level prediction of the forward-backward asymmetry as a function of $e^+e^-$ invariant mass for $u\bar u \to e^+e^-$ and $d\bar d \to e^+e^-$.

With sufficient statistics in Run II the forward-backward asymmetry can be used to measure $\sin^2\bar\theta_W$, which in turn can provide a constraint on the standard model complementary to the measurement of the $W$ boson mass. The Tevatron also allows measurement of the asymmetry at partonic center-of-mass energies above the center of mass energy of LEP II. This measurement can be used, not only to confirm the standard model $\gamma^*/Z$ interference which dominates in this region, but also to investigate possible new phenomena which may alter $A_{FB}$, such as new neutral gauge bosons [4] or large extra dimensions [5].

CDF have measured the forward-backward asymmetry at the Tevatron using $e^+e^-$ pairs in 110 pb$^{-1}$ of data at $\sqrt{s} = 1.8$ TeV [6]. They obtain $A_{FB} = 0.070 \pm 0.016$ in the mass region 75 GeV $< m_{e^+e^-} <$ 105 GeV, and $A_{FB} = 0.43 \pm 0.10$ in the region $m_{e^+e^-} >$ 105 GeV. The much larger Run II statistics will enable $A_{FB}$ to be measured with an uncertainty reduced by over an order of magnitude.

## 2. Simulation

The simulations presented here use the ZGRAD Monte Carlo program [7], which includes $\mathcal{O}(\alpha)$ QED radiative corrections to the process $p\bar{p} \to \gamma^*/Z \to \ell^+\ell^-$. We simulate this process at $\sqrt{s} = 2.0$ TeV using the MRST parton distributions as our default set. Since the radiative corrections are included in ZGRAD, we denote the process of interest by $p\bar{p} \to \gamma^*/Z \to \ell^+\ell^-(\gamma)$ in the remainder of this paper. The ZGRAD program includes real and virtual corrections in the initial and final states.

In our simulations, the effects of detector resolution are modeled by smearing the 4-momenta of the particles from ZGRAD according to the estimated resolution of the Run II DØ detector. We smear the 4-momenta of electrons, positrons and photons according to the energy resolution $\sigma_{EM}$ of the calorimeters, which have been parametrized using constant, sampling and noise terms as

$$\left(\frac{\sigma_{EM}}{E}\right)^2 = \begin{cases} c_{EM}^2 + \left(\frac{s_{EM}}{\sqrt{E_T}}\right)^2 + \left(\frac{n_{EM}}{E}\right)^2 \\ \qquad\qquad Central\ Calorimeter \\ c_{EM}^2 + \left(\frac{s_{EM}}{\sqrt{E}}\right)^2 + \left(\frac{n_{EM}}{E}\right)^2 \\ \qquad\qquad End\ Calorimeters \end{cases} \quad (1)$$

where we use the parameters relevant for the Run I detector, $c_{EM} = 0.0115$, $s_{EM} = 0.135$, and $n_{EM} = 0.43$ for the CC, and $c_{EM} = 0.0100$, $s_{EM} = 0.157$, and $n_{EM} = 0.29$ for the EC. With the addition of the 2 T solenoidal magnetic field in Run II, only minor changes in these parameters are expected. The transverse momentum of muons in the Run II detector will be measured in the central tracking system, consisting of the Central Fiber Tracker (CFT) and the Silicon Microstrip Tracker (SMT). The momentum resolution of the tracking system has been studied using the fast Monte Carlo MCFAST. From these studies the resolution in $1/p_T$ is parametrized as:

$$\sigma\left(\frac{1}{p_T}\right) = \sqrt{\left(\frac{\alpha}{L^2}\right)^2 + \left(\frac{\gamma}{p_T\sqrt{L|\sin\theta|}}\right)^2} \quad (2)$$

where

$$L = \begin{cases} 1 & 0 < \theta \leq \theta_c \\ \frac{\tan\theta}{\tan\theta_c} & \theta_c < \theta < 90° \end{cases}. \quad (3)$$

Here $\alpha = 0.0017$ GeV$^{-1}$, $\gamma = 0.018$, $L$ is the fraction of the projection of the track length in the bending plane which is measured in the Tracker, and $\theta_c \approx 23°$ is the polar angle beyond which the number of CFT layers crossed by a track starts to decrease. The first term in Eq. (3) is due to the detector resolution while the second term is due to multiple scattering.

Figure 2 shows the transverse momentum resolution as a function of detector pseudorapidity $|\eta_{det}|$ for tracks with a $p_T$ of 1, 20 and 100 GeV, while Fig. 3 shows the resolution as a function of $p_T$ and $|\eta_{det}|$ in the form of a contour plot. For central tracks ($\eta = 0$) with $p_T = 45$ GeV, the resolution is $\sigma(p_T)/p_T = 8\%$, to be compared with the calorimeter energy resolution for 45 GeV electrons of $\sigma(E)/E = 2.5\%$.

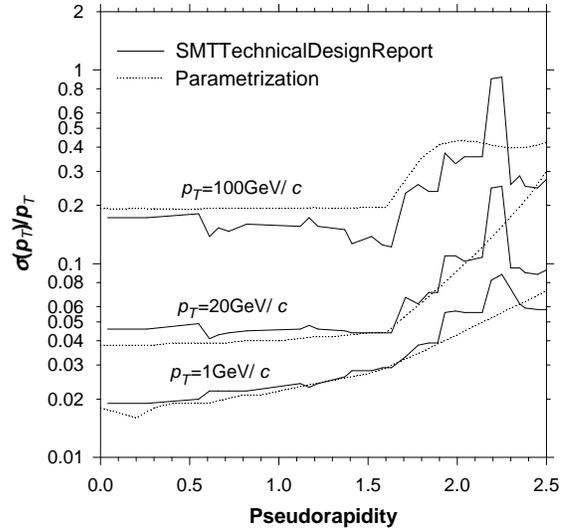

Figure 2. Parametrized transverse momentum resolution for the DØ Run II tracking system (dotted lines). The solid lines are the results of a simple Monte Carlo simulation taken from the DØ SMT Technical Design Report [8].

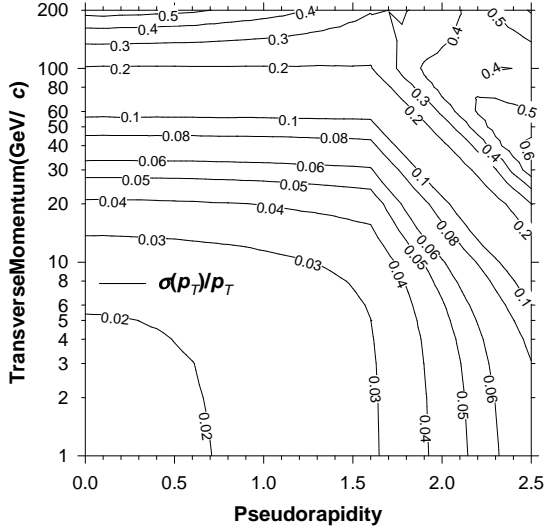

Figure 3. Transverse momentum resolution for the central tracking system plotted as a function of transverse momentum and pseudorapidity.

We assume an overall detection efficiency of 75% for $e^+e^-$ events and 65% for $\mu^+\mu^-$ events. These efficiencies are rough estimates of the effects of trigger and particle identification efficiencies expected in Run II. The results can be updated once more realistic numbers for these efficiencies become available.

The ZGRAD program generates weighted events. Due to the occurrence of negative weights, we did not attempt to unweight the events. Thus, we work with weighted events and properly account for the weights in our calculations of the forward-backward asymmetry errors. The forward-backward asymmetry is defined by

$$A_{FB} = \frac{\sigma_F - \sigma_B}{\sigma_F + \sigma_B} \qquad (4)$$

where $\sigma_F$ and $\sigma_B$ are the forward and backward cross sections, defined by

$$\sigma_F = \int_0^1 \frac{d\sigma}{d(\cos\theta^*)} d(\cos\theta^*)$$
$$\sigma_B = \int_{-1}^0 \frac{d\sigma}{d(\cos\theta^*)} d(\cos\theta^*) \qquad (5)$$

and $\theta^*$ is the angle of the lepton in the Collins-Soper frame [9].

The statistical error on $A_{FB}$ is given by

$$\delta A_{FB} = 2 \frac{\sqrt{\sigma_B^2 (\delta\sigma_F)^2 + \sigma_F^2 (\delta\sigma_B)^2}}{(\sigma_F + \sigma_B)^2} \qquad (6)$$

where $\delta\sigma_F$, $\delta\sigma_B$ are the uncertainties in the forward and backward cross sections. For unweighted events, this simplifies to

$$\delta A_{FB} = \frac{2}{N_F + N_B} \sqrt{\frac{N_F N_B}{N_F + N_B}} \qquad (7)$$

where $N_F$, $N_B$ are the numbers of forward and backward events. However, ZGRAD generates weighted events and, therefore, we use Eq. (6) where $\delta\sigma_F$ and $\delta\sigma_B$, are calculated using the appropriate event weights.

The selection cuts used in our study are summarized in Table 1. In the electron channel we require one of the electrons to be in the CC ($|\eta_{det}| < 1.0$), while the other electron may be in the CC or in the EC ($|\eta_{det}| < 1.0$ or $1.5 < |\eta_{det}| < 2.5$).

In the muon channel we require both muons to be within $|\eta_{det}| < 1.7$. In Run II the muon coverage is expected to extend up to $|\eta_{det}| = 2.0$. We chose to limit the muon acceptance to $|\eta_{det}| = 1.7$ since Monte Carlo events were already generated with this restriction and large CPU time would have been required to re-generate the events.

We account in our simulation for the granularity of the DØ calorimeter. If the photon is very close to the electron its energy will be merged with that of the electron cluster. Thus, in the simulation we combine the photon and electron 4-momenta to form an effective electron 4-momentum if the photon is within $\Delta R_{e\gamma} \equiv \sqrt{\Delta\eta_{e\gamma}^2 + \Delta\phi_{e\gamma}^2} < 0.2$. If the photon falls within $0.2 < \Delta R_{e\gamma} < 0.4$, we reject the event if $E_\gamma/(E_e + E_\gamma) > 0.15$, since the event will not pass the standard isolation criterion imposed on electrons.

If a photon is very close to a muon and it deposits sufficient energy in the calorimeter close to the muon track, the energy deposition in the calorimeter will not be consistent with the passage of a minimum ionizing muon. Therefore, in the simulation we reject events if $\Delta R_{\mu\gamma} < 0.2$ and $E_\gamma > 2$ GeV.

Table 1
Selection criteria for $e^+e^-$ and $\mu^+\mu^-$ events.

|  | $e^+e^-$ | | $\mu^+\mu^-$ | |
| --- | --- | --- | --- | --- |
| Selection cut | $e_1$ | $e_2$ | $\mu_1$ | $\mu_2$ |
| $p_T$ (GeV) | > 25 | > 25 | > 20 | > 15 |
| $|\eta_{det}|$ | < 1.0 | < 1.0 or 1.5 − 2.5 | < 1.7 | < 1.7 |
| $m_{\ell^+\ell^-}$ (GeV) | > 40 | > 40 | > 40 | > 40 |

## 3. Results

The $\ell^+\ell^-$ invariant mass distributions for $p\bar{p} \to \gamma^*/Z \to \ell^+\ell^-(\gamma)$ at $\sqrt{s} = 2.0$ TeV from the ZGRAD simulations, using the MRST parton distribution functions are shown in Fig. 4. The thin line shows $d\sigma/dm_{\ell^+\ell^-}$ without any kinematic cuts applied and with no detector acceptance or resolution effects included. In order to obtain sufficient statistical precision a large number of events were generated in multiple runs covering overlapping regions of $m_{\ell^+\ell^-}$. The thick line shows $d\sigma/dm_{\ell^+\ell^-}$ after kinematic cuts and detector effects are included. The error bars represent the statistical errors only, calculated from Eq. (6), assuming an integrated luminosity of 10 fb$^{-1}$.

Fig. 5 shows the forward-backward asymmetry as a function of $m_{\ell^+\ell^-}$. The solid line shows $A_{FB}$ without any kinematic cuts applied and with no detector acceptance or resolution effects included. The solid points show $A_{FB}$ after kinematic cuts and detector effects are included. The error bars represent the statistical errors only, calculated from Eq. (6), assuming an integrated luminosity of 10 fb$^{-1}$. As can be seen, detector resolution and acceptance effects significantly alter the shape of the $A_{FB}$ vs. $m_{\ell^+\ell^-}$ curve, especially at low di-lepton invariant masses. In this region, the effect of CC/EC acceptance increases $A_{FB}$, while restricting $\mu^+\mu^-$ events to be in the central region decreases the asymmetry. This is also true of $e^+e^-$ events if only CC/CC events are considered. In the vicinity of the Z-pole the energy resolution is better than the $p_T$ resolution, and hence the $A_{FB}^{e^+e^-}$ is altered less than $A_{FB}^{\mu^+\mu^-}$. In these plots the $A_{FB}$ shown is the reconstructed $A_{FB}$ without corrections for acceptance or resolution effects.

In order to obtain a measurement of the weak mixing angle we assume the relationship

$$A_{FB} = a + b\,\sin^2\bar{\theta}_W \qquad (8)$$

so that the statistical error on $\sin^2\bar{\theta}_W$ is given by

$$\delta \sin^2 \bar{\theta}_W = \frac{\delta A_{FB}}{b}. \qquad (9)$$

The quantity $b$ is determined by varying $\sin^2\bar{\theta}_W$ in the Monte Carlo simulations. Since $A_{FB}$ is determined over a finite range of di-lepton invariant mass, we have investigated the effect of the lower and upper $m_{\ell^+\ell^-}$ cuts on $\delta \sin^2 \bar{\theta}_W$. The optimal precision is obtained for 75 GeV $< m_{\ell^+\ell^-} <$ 105 GeV, i.e. a mass window encompassing the Z-pole. This is to be expected because the sensitivity $b$ is maximal at the Z-pole and this region is where the cross section peaks and hence the statistical error is smallest. Thus, the $A_{FB}$ values and errors presented in the remainder of this paper are all obtained with a di-lepton invariant mass cut of 75 GeV $< m_{\ell^+\ell^-} <$ 105 GeV. Table 2 shows the resulting statistical uncertainties obtained from the electron and muon channels. The $e^+e^-$ and $\mu^+\mu^-$ channels yield similar uncertainties on $A_{FB}$ and $\sin^2 \bar{\theta}_W$. In both channels the effect of the selection cuts is to reduce the sensitivity $b$ from about 3.5 to about 2.8.

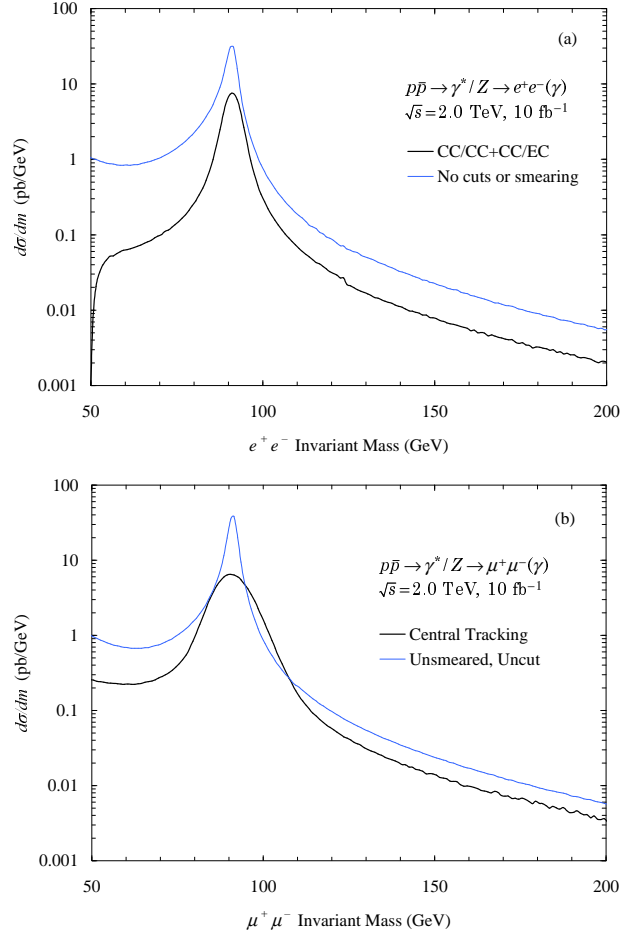

Figure 4. Invariant mass $m_{\ell^+\ell^-}$ distributions for (a) $e^+e^-$ events and (b) $\mu^+\mu^-$ events. The thin line is the distribution obtained with no cuts or detector effects applied and the thick line is the resulting distribution after selection cuts and detector effects are included.

The effects of NLO QCD corrections to the process $p\bar{p} \to \gamma^*/Z \to \ell^+\ell^-$ are not included in ZGRAD, so we estimate these using the $\mathcal{O}(\alpha_s)$ event generator described in [7]. Using the same method as described in Section 2, we calculate the change in the forward-backward asymmetry and the shift in sensitivity due

Table 2
Uncertainties on $A_{FB}$ and $\sin^2\bar\theta_W$ in the invariant mass range 75 GeV $< m_{\ell^+\ell^-} <$ 105 GeV for an integrated luminosity of 10 fb$^{-1}$. Also shown are the assumed event detection efficiency, the number of events passing all the cuts $N_{obs}$, the forward-backward asymmetry $A_{FB}$, and the sensitivity $b$.

| Process | Selection cuts | Efficiency | $N_{obs}$ | $A_{FB}$ | $\delta A_{FB}^{stat}$ | $b$ | $\delta \sin^2\bar\theta_W$ |
|---|---|---|---|---|---|---|---|
| $Z \to e^+e^-$ | no | 100% | $1.78 \times 10^6$ | 0.0551 | 0.0008 | 3.43 | 0.0002 |
| $Z \to e^+e^-$ | yes | 75% | $3.82 \times 10^5$ | 0.0515 | 0.0014 | 2.78 | 0.0005 |
| $Z \to \mu^+\mu^-$ | no | 100% | $1.87 \times 10^6$ | 0.0534 | 0.0007 | 3.51 | 0.0002 |
| $Z \to \mu^+\mu^-$ | yes | 65% | $5.67 \times 10^5$ | 0.0420 | 0.0011 | 2.62 | 0.0004 |

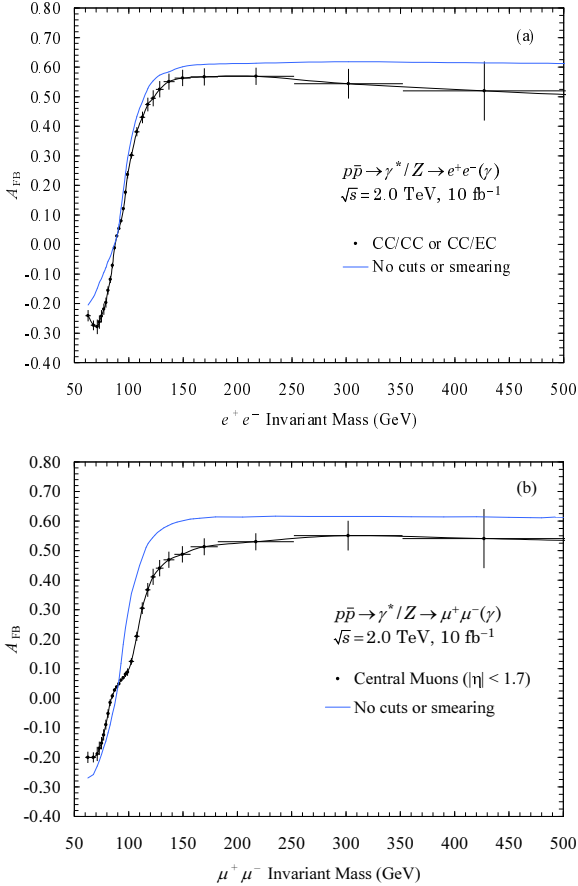

Figure 5. Forward-backward asymmetry $A_{FB}$ vs. di-lepton invariant mass for (a) $e^+e^-$ events and (b) $\mu^+\mu^-$ events. The solid line is the distribution obtained with no cuts or detector effects applied and the points are the resulting distribution after selection cuts and detector effects are included. The error bars represent the statistical errors for a data sample of 10 fb$^{-1}$.

to NLO QCD corrections. Thus, we write

$$\Delta A_{FB} = A_{FB}^{\mathcal{O}(\alpha_s)} - A_{FB}^{LO} \qquad (10)$$
$$\Delta b = b^{\mathcal{O}(\alpha_s)} - b^{LO} \qquad (11)$$

where $LO$ denotes the leading-order quantities. For events generated including detector effects we find the shift in $A_{FB}$ to be negligible for $e^+e^-$ events and $-13\%$ for $\mu^+\mu^-$ events. The shift in sensitivity is $\Delta b/b^{LO} \approx -3.4\%$ for $e^+e^-$ events and $\approx -25\%$ for $(\mu^+\mu^-)$ events. Thus, NLO QCD effects decrease the sensitivity to $\sin^2\bar\theta_W$ by 3.4% in the $e^+e^-$ channel and 25% in the $\mu^+\mu^-$ channel.

## 4. Systematic Uncertainties

### 4.1. Parton Distribution Functions

Since the vector and axial couplings of the $u$ and $d$ quarks to the $Z$ boson are different, the lepton forward-backward asymmetry is expected to depend on the ratio of the $u$ to $d$ quark parton distribution functions. Thus, the choice of the parton distribution functions (PDF's) will affect the measured lepton forward-backward asymmetry.

We have run simulations with six PDF's from the MRS [ 10] and CTEQ [ 11] sets to study the effect of the PDF's on the asymmetry. Fig. 6 shows the $e^+e^-$ and $\mu^+\mu^-$ asymmetries and their statistical errors for each PDF.

The largest deviation from the MRST value for $A_{FB}$ is 0.0018 for the $e^+e^-$ channel and 0.0015 for the $\mu^+\mu^-$ channel. While these numbers are of the same order as the statistical error expected on $A_{FB}$ for 10 fb$^{-1}$, we expect that in Run II our knowledge of the PDF's will improve considerably, e.g. from the constraints imposed by the Run II $W$ asymmetry measurements. Thus, we expect a significantly decreased systematic error due to the uncertainty in the PDF's which will likely render it insignificant compared with the statistical error in the measurement. For example, if the PDF uncertainty scales as $1/\sqrt{N}$, the uncertainty in $A_{FB}$ would be $\delta A_{FB} \approx 0.00018$ (0.00015) for an integrated luminosity of 10 fb$^{-1}$ in the $e^+e^-$ ($\mu^+\mu^-$) channel.

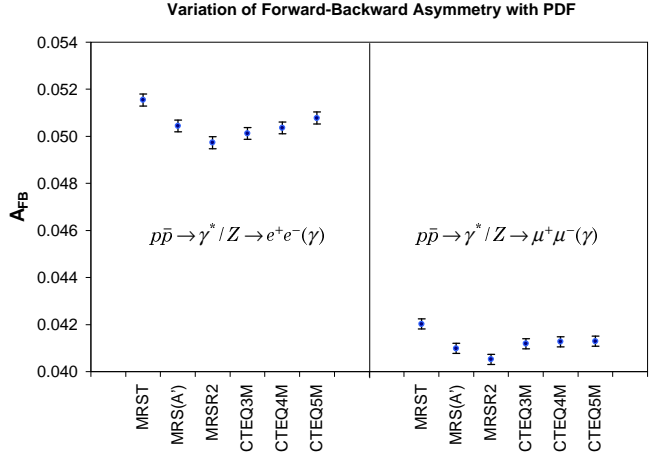

Figure 6. Variation of the forward-backward asymmetry for $e^+e^-$ and $\mu^+\mu^-$ events, including selection cuts and detector effects.

### 4.2. Energy Scale Calibration Uncertainties

The energy scale uncertainty, or the uncertainty in mapping the calorimeter response to the true electron energy, affects the forward-backward asymmetry by causing a shift in the $\ell^+\ell^-$ invariant mass range over which we integrate $A_{FB}$. The effect is significant in the Z-pole region, but at large invariant masses $A_{FB}$ is essentially constant and the energy scale uncertainty does not play a role. For electrons, the measured energy can be related to the true energy by

$$E_{meas} = \alpha E_{true} + \delta \qquad (12)$$

where the scale factor $\alpha$ and offset $\delta$ are determined by calibration of the calorimeters. In Run I DØ determined $\alpha = 0.9533 \pm 0.0008$ and $\delta = -0.16^{+0.03}_{-0.21}$ GeV.

Assuming these uncertainties in $\alpha$ and $\delta$, we find a systematic error of $\delta A_{FB} = 0.0002$ due to the overall energy scale uncertainty.

### 4.3. Uncertainty due to Backgrounds

Backgrounds are not included in the simulations above, but we can estimate the uncertainty due to backgrounds as follows. If we assume that the fraction of observed events which are due to backgrounds is $\alpha \pm \delta\alpha$, then the uncertainty in the forward-backward asymmetry will be

$$\delta A_{FB} = 2\sqrt{2} \frac{N_F N_B}{(N_F + N_B)^2} \delta\alpha \qquad (13)$$

where $N_F$, $N_B$ are the numbers of background-subtracted forward and backward events, and we have assumed that the background events are symmetric in $\cos\theta^*$. If we assume that the error in the background fraction $\delta\alpha$ scales as the inverse of the integrated luminosity, we can extrapolate from the uncertainties in the Run I CDF and DØ data samples to estimate the error. The Run I uncertainties were $\delta\alpha \approx 0.1 - 1.0\%$. Thus, for an integrated luminosity of 10 fb$^{-1}$, we obtain an uncertainty on $A_{FB}$ of $\delta A_{FB} = 0.00014$.

### 4.4. Summary of Uncertainties in $\sin^2\bar{\theta}_W$

Table 3 summarizes the statistical and individual systematic uncertainty estimates expected with 10 fb$^{-1}$ of data. We estimated the uncertainties due to electron energy resolution and muon transverse momentum resolution to be negligible.

### 4.5. Conclusions

The measurement of the forward-backward asymmetry in $e^+e^-$ and $\mu^+\mu^-$ events in Run II provides a means to test the standard model $\gamma^*/Z$ interference at $\ell^+\ell^-$ invariant masses well above the center of mass energy of the LEP II collider. The estimated DØ precision on $A_{FB}$ achievable with 10 fb$^{-1}$ integrated luminosity is shown in Fig. 5.

In the vicinity of the Z-pole this measurement can also be used to determine the effective weak mixing angle $\sin^2\bar{\theta}_W$. The optimal precision on $\sin^2\bar{\theta}_W$ is obtained for 75 GeV $< m_{\ell^+\ell^-} <$ 105 GeV, i.e. a mass window encompassing the Z-pole. This is to be expected because the sensitivity $b$ is maximal at the Z-pole and this region is where the cross section peaks and hence the statistical error is smallest. For 10 fb$^{-1}$ we estimate that the total error on $\sin^2\bar{\theta}_W$ will be 0.0005 in the electron channel and 0.0006 in the muon channel, assuming that systematic errors scale as the inverse of the square root of the integrated luminosity. One would expect similar precision from CDF, and combining the results of the two experiments in both channels the overall uncertainty would be $\delta\sin^2\bar{\theta}_W \approx 0.00028$. Therefore, if integrated luminosities in excess of 10 fb$^{-1}$ can be achieved in Run II, it appears that the determination of $\sin^2\bar{\theta}_W$ will have comparable precision to the current world average of the measurements from LEP and SLD.

## 5. Measuring $A_{FB}$ at the LHC

At the LHC, the $Z \to \ell^+\ell^-$ cross section is approximately a factor 7 larger than at the Tevatron. However, the measurement of the forward backward asymmetry is complicated by several factors. In $pp$ collisions, the quark direction in the initial state has to be extracted from the boost direction of the $\ell^+\ell^-$ system with respect to the beam axis. At LHC energies, the sea-sea quark flux is much larger than at the Tevatron. As a result, the probability, $f_q$, that the quark direction and the boost direction of the di-lepton system coincide is rather small. The forward backward asymmetry is therefore smaller than at the Tevatron, and the sensitivity to $\sin^2\bar\theta_W$ at the LHC with 100 fb$^{-1}$ per lepton channel and experiment [ 7, 12] is similar to that estimated for the Tevatron with 10 fb$^{-1}$ (see Sec. 4.5).

Table 3
Summary of uncertainties on $A_{FB}$ and $\sin^2\bar\theta_W$ in the invariant mass range 75 GeV $< m_{\ell^+\ell^-} <$ 105 GeV for an integrated luminosity of 10 fb$^{-1}$. The effects of $\mathcal{O}(\alpha)$ QED corrections and NLO QCD corrections have been taken into account.

|  | $e^+e^-$ | | $\mu^+\mu^-$ | |
| --- | --- | --- | --- | --- |
| Error source | $\delta A_{FB}$ | $\delta\sin^2\bar\theta_W$ | $\delta A_{FB}$ | $\delta\sin^2\bar\theta_W$ |
| Statistical | 0.0014 | 0.0005 | 0.0011 | 0.0006 |
| Systematics | 0.0003 | 0.0001 | 0.0002 | 0.00010 |
| PDF | 0.00018 | 0.00007 | 0.00015 | 0.00008 |
| EM scale | 0.0002 | 0.00007 | – | – |
| Backgrounds | 0.00014 | 0.00005 | 0.00014 | 0.00007 |
| Total uncertainty | 0.0014 | 0.0005 | 0.0011 | 0.0006 |

Restricting the $A_{FB}$ measurement to events which satisfy $|y(\ell^+\ell^-)| > 1$ in addition to the $|y(\ell)| < 2.5$ cut improves the significance of the measurement by about a factor 1.5. Events with a large di-lepton rapidity originate from collisions where at least one of the partons carries a large fraction $x$ of the proton momentum. Since valence quarks dominate at high values of $x$, a cut on $y(\ell^+\ell^-)$ increases $f_q$ and thus the asymmetry. However, the gain due to the larger asymmetry is partially cancelled by the loss of statistics, leaving a modest improvement only.

In order to achieve a precision better than the current value of $\delta\sin^2\bar\theta_W = 1.7 \times 10^{-4}$ [ 13], it will be necessary to detect one of the leptons in the rapidity range up to $|y(\ell)| < 5$ at the LHC. If this can be done, one expects that the weak mixing angle can be determined with a precision of

$$\delta\sin^2\bar\theta_W = 1.4 \times 10^{-4}, \qquad (14)$$

per lepton channel and experiment for an integrated luminosity of 100 fb$^{-1}$. In order to reach the precision given in Eq. (14), a jet rejection factor of $10 - 100$ has to be achieved in the forward rapidity region $2.5 < |y(\ell)| < 5$, and the lepton acceptance times the reconstruction efficiency as a function of $y(\ell)$ has to be known to 0.1% or better [ 12].

For comparison, at a Linear Collider operating at $\sqrt{s} = M_Z$ with a luminosity of a few $\times 10^{33}$ cm$^{-2}$ s$^{-1}$, it is expected that the weak mixing angle be determined with a precision of about $\delta\sin^2\bar\theta_W = 1 \times 10^{-5}$ [ 14].

# Global Fits to Electroweak Data Using GAPP

J. Erler[a]

[a]Department of Physics and Astronomy, University of Pennsylvania,
Philadelphia, PA 19104-6396

At Run II of the Tevatron it will be possible to measure the $W$ boson mass with a relative precision of about $2 \times 10^{-4}$, which will eventually represent the best measured observable beyond the input parameters of the SM. Proper interpretation of such an ultra-high precision measurement, either within the SM or beyond, requires the meticulous implementation and control of higher order radiative corrections. The FORTRAN package GAPP, described here, is specifically designed to meet this need and to ensure the highest possible degrees of accuracy, reliability, adaptability, and efficiency.

## 1. PRECISION TESTS

Precision analysis of electroweak interactions follows three major objectives: high precision tests of the SM; the determination of its fundamental parameters; and studies of indications and constraints of possible new physics beyond the SM, such as supersymmetry or new gauge bosons. Currently, the experimental information comes from the very high precision $Z$ boson measurements at LEP and the SLC, direct mass measurements and constraints from the Tevatron and LEP II, and low energy precision experiments, such as in atomic parity violation, $\nu$ scattering, and rare decays. These measurements are compared with the predictions of the SM and its extensions. The level of precision is generally very high. Besides the need for high-order loop calculations, it is important to utilize efficient renormalization schemes and scales to ensure sufficient convergence of the perturbative expansions.

The tasks involved called for the creation of a special purpose FORTRAN package, GAPP, short for the *Global Analysis of Particle Properties* [ 1]. It is mainly devoted to the calculation of pseudo-observables, i.e., observables appropriately idealized from the experimental reality. The reduction of raw data to pseudo-observables is performed by the experimenters with available packages (e.g., ZFITTER for $Z$ pole physics). For cross section and asymmetry measurements at LEP II (not implemented in the current version, GAPP_99.7), however, this reduction is not optimal and convoluted expressions should be used instead. GAPP attempts to gather all available theoretical and experimental information; it allows the addition of extra parameters describing new physics; it treats all relevant SM inputs as global fit parameters; and it can easily be updated with new calculations, data, observables, or fit parameters. For clarity and speed it avoids numerical integrations throughout. It is based on the modified minimal subtraction ($\overline{MS}$) scheme which demonstrably avoids large expansion coefficients.

GAPP is endowed with the option to constrain nonstandard contributions to the *oblique parameters* defined to affect only the gauge boson self-energies [ 2] (e.g. $S$, $T$, and $U$); specific anomalous $Z$ couplings; the number of active neutrinos (with standard couplings to the $Z$ boson); and the masses, mixings, and coupling strengths of extra $Z$ bosons appearing in models of new physics. With view on the importance of supersymmetric extensions of the SM on one hand, and upcoming experiments on the other, I also included the $b \to s\gamma$ transition amplitude, and intend to add the muon anomalous magnetic moment. In the latter case, there are theoretical uncertainties from hadronic contributions which are partially correlated with the renormalization group (RG) evolutions of the QED coupling and the weak mixing angle. These correlations will be partially taken into account by including heavy quark effects in analytical form; see Ref. [ 3] for a first step in this direction. By comparing this scheme with more conventional ones, it will also be possible to isolate a QCD sum rule and to rigorously determine the charm and bottom quark $\overline{MS}$ masses, $\hat{m}_c$ and $\hat{m}_b$, with high precision.

## 2. GAPP

### 2.1. Basic structure

In the default running mode of the current version, GAPP_99.7, a fit is performed to 41 observables, out of which 26 are from $Z$ pole measurements at LEP and the SLC. The Fermi constant, $G_F$ (from the muon lifetime), the electromagnetic fine structure constant, $\alpha$ (from the quantum Hall effect), and the light fermion masses are treated as fixed inputs. The exception is $\hat{m}_c$ which strongly affects the RG running[1] of $\hat{\alpha}(\mu)$ for $\mu > \hat{m}_c$. I therefore treat $\hat{m}_c$ as a fit parameter and include an external constraint with an enhanced error to absorb hadronic threshold uncertainties of other quark flavors, as well as theoretical uncertainties from the application of perturbative QCD at relatively low energies. Other fit parameters are the $Z$ boson mass, $M_Z$, the Higgs boson mass, $M_H$, the top quark mass, $m_t$, and the strong coupling constant, $\alpha_s$, so that there are 37 effective degrees of freedom. Given current precisions, $M_Z$ may alternatively be treated as an additional fixed input.

---
[1]Quantities defined in the $\overline{MS}$ scheme are denoted by a caret.

The file `fit.f` basically contains a simple call to the minimization program MINUIT [4] (from the CERN program library) which is currently used in data driven mode (see `smfit.dat`). It in turn calls the core subroutine `fcn` and the $\chi^2$-function `chi2`, both contained in `chi2.f`. Subroutine `fcn` defines constants and flags; initializes parts of the one-loop package FF [5, 6]; and makes the final call to subroutine `values` in `main.f` which drives the output (written to file `smfit.out`). In `chi2` the user actively changes and updates the data for the central values, errors, and correlation coefficients of the observables, and includes or excludes individual contributions to $\chi^2$ (right after the initialization, `chi2 = 0.d0`). To each observable (as defined at the beginning of `chi2`) corresponds an entry in each of the fields `value, error, smval`, and `pull`, containing the central observed value, the total (experimental and theoretical) error, the calculated fit value, and the standard deviation, respectively. The function `chi2` also contains calls to various other subroutines where the actual observable calculations take place. These are detailed in the following subsections.

Another entry to GAPP is provided through `mh.f` which computes the probability distribution function of $M_H$. The probability distribution function is the quantity of interest within Bayesian data analysis (as opposed to point estimates frequently used in the context of classical methods), and defined as the product of a prior density and the likelihood, $\mathcal{L} \sim \exp(-\chi^2/2)$. If one chooses to disregard any further information on $M_H$ (such as from triviality considerations or direct searches) one needs a *non-informative prior*. It is recommended to choose a flat prior in a variable defined on the whole real axis, which in the case of $M_H$ is achieved by an equidistant scan over $\log M_H$. An *informative prior* is obtained by activating one of the approximate Higgs exclusion curves from LEP II near the end of `chi2.f`. These curves affect values of $M_H$ even larger than the corresponding quoted 95% CL lower limit and includes an extrapolation to the kinematic limit; notice that this corresponds to a conservative treatment of the upper $M_H$ limit.

Contour plots can be obtained using the routine `mncontours` from MINUIT. For the cases this fails, some simpler and slower but more robust contour programs are also included in GAPP, but these have to be adapted by the user to the case at hand.

**2.2.** $\hat{\alpha}$, $\sin^2 \hat{\theta}_W$, $M_W$

At the core of present day electroweak analyses is the interdependence between $G_F$, $M_Z$, the $W$ boson mass, $M_W$, and the weak mixing angle, $\sin^2 \theta_W$. In the $\overline{\text{MS}}$ scheme it can be written as [7, 8],

$$\hat{s}^2 = \frac{A^2}{M_W^2(1 - \Delta\hat{r}_W)}, \qquad \hat{s}^2\hat{c}^2 = \frac{A^2}{M_Z^2(1 - \Delta\hat{r}_Z)}, \quad (1)$$

where,

$$A = \left[\frac{\pi\alpha}{\sqrt{2}G_F}\right]^{1/2} = 37.2805(2) \text{ GeV}, \quad (2)$$

$\hat{s}^2$ is the $\overline{\text{MS}}$ mixing angle, $\hat{c}^2 = 1 - \hat{s}^2$, and where,

$$\Delta\hat{r}_W = \frac{\alpha}{\pi}\hat{\Delta}_\gamma + \frac{\hat{\Pi}_{WW}(M_W^2) - \hat{\Pi}_{WW}(0)}{M_W^2} + V + B, \quad (3)$$

and,

$$\Delta\hat{r}_Z = \Delta\hat{r}_W + (1 - \Delta\hat{r}_W)\frac{\hat{\Pi}_{ZZ}(M_Z^2) - \frac{\hat{\Pi}_{WW}(M_W^2)}{\hat{c}^2}}{M_Z^2}, \quad (4)$$

collect the radiative corrections computed in `sin2th.f`. The $\hat{\Pi}$ indicate $\overline{\text{MS}}$ subtracted self-energies, and V + B denote the vertex and box contributions to $\mu$ decay. Although these relations involve the $\overline{\text{MS}}$ gauge couplings they employ on-shell gauge boson masses, absorbing a large class of radiative corrections [9].

$\Delta\hat{r}_W$ and $\Delta\hat{r}_Z$ are both dominated by the contribution $\hat{\Delta}_\gamma(M_Z)$ which is familiar from the RG running of the electromagnetic coupling,

$$\hat{\alpha}(\mu) = \frac{\alpha}{1 - \frac{\alpha}{\pi}\hat{\Delta}_\gamma(\mu)}, \quad (5)$$

and computed in `alfahat.f` up to four-loop $\mathcal{O}(\alpha\alpha_s^3)$. Contributions from $c$ and $b$ quarks are calculated using an unsubtracted dispersion relation [3]. If $\mu$ is equal to the mass of a quark, three-loop matching is performed and the definition of $\hat{\alpha}$ changes accordingly. Pure QED effects are included up to next-to-leading order (NLO) while higher orders are negligible. Precise results can be obtained for $\mu < 2m_\pi$ and $\mu \geq m_c$.

Besides full one-loop electroweak corrections, $\Delta\hat{r}_W$ and $\Delta\hat{r}_Z$ include enhanced two-loop contributions of $\mathcal{O}(\alpha^2 m_t^4)$ [10] (implemented using the analytic expressions of Ref. [11]) and $\mathcal{O}(\alpha^2 m_t^2)$ [12] (available as expansions in small and large $M_H$); mixed electroweak/QCD corrections of $\mathcal{O}(\alpha\alpha_s)$ [13] and $\mathcal{O}(\alpha\alpha_s^2 m_t^2)$ [14]; the analogous mixed electroweak/QED corrections of $\mathcal{O}(\alpha^2)$; and fermion mass corrections also including the leading gluonic and photonic corrections.

## 2.3. Z decay widths and asymmetries

The partial width for $Z \to f\bar{f}$ decays is given by,

$$\Gamma_{f\bar{f}} = \frac{N_C^f M_Z \hat{\alpha}}{24 \hat{s}^2 \hat{c}^2} |\hat{\rho}_f| \left[1 - 4|Q_f|\text{Re}(\hat{\kappa}_f)\hat{s}^2 + 8Q_f^2 \hat{s}^4 |\hat{\kappa}_f|^2\right]$$

$$\times \left[1 + \delta_{\text{QED}} + \delta_{\text{QCD}}^{\text{NS}} + \delta_{\text{QCD}}^{\text{S}} - \frac{\hat{\alpha}\hat{\alpha}_s}{4\pi^2} Q_f^2 + \mathcal{O}(m_f^2)\right]. \quad (6)$$

$N_C^f$ is the color factor, $Q_f$ is the fermion charge, and $\hat{\rho}_f$ and $\hat{\kappa}_f$ are form factors which differ from unity through one-loop electroweak corrections [15] and are computed in rho.f and kappa.f, respectively. For $f \neq b$ there are no corrections of $\mathcal{O}(\alpha^2 m_t^4)$ and contributions of $\mathcal{O}(\alpha^2 m_t^2)$ to $\hat{\kappa}_f$ [16] and $\hat{\rho}_f$ [9] are very small and presently neglected. On the other hand, vertex corrections of $\mathcal{O}(\alpha\alpha_s)$ [17] are important and shift the extracted $\alpha_s$ by $\sim 0.0007$.

The $Z \to b\bar{b}$ vertex receives extra corrections due to heavy top quark loops. They are large and have been implemented in bvertex.f based on Ref. [18]. $\mathcal{O}(\alpha^2 m_t^4)$ corrections [10, 11] are included, as well, while those of $\mathcal{O}(\alpha^2 m_t^2)$ are presently unknown. The leading QCD effects of $\mathcal{O}(\alpha\alpha_s m_t^2)$ [19] and all subleading $\mathcal{O}(\alpha\alpha_s)$ corrections [20] are incorporated into $\hat{\rho}_b$ and $\hat{\kappa}_b$, but not the $\mathcal{O}(\alpha\alpha_s^2 m_t^2)$ contribution which is presently available only for nonsinglet diagrams [21].

In Eq. (6), $\delta_{\text{QED}}$ are the $\mathcal{O}(\alpha)$ and $\mathcal{O}(\alpha^2)$ QED corrections. $\delta_{\text{QCD}}^{\text{NS}}$ are the universal QCD corrections up to $\mathcal{O}(\alpha_s^3)$ which include quark mass dependent contributions due to double-bubble type diagrams [22, 23]. $\delta_{\text{QCD}}^{\text{S}}$ are the singlet contributions to the axial-vector and vector partial widths which start, respectively, at $\mathcal{O}(\alpha_s^2)$ and $\mathcal{O}(\alpha_s^3)$, and induce relatively large family universal but flavor non-universal $m_t$ effects [23, 24]. The corrections appearing in the second line of Eq. (6) are evaluated in lep100.f.

The dominant massless contribution to $\delta_{\text{QCD}}^{\text{NS}}$ can be obtained by analytical continuation of the Adler $D$-function, which (in the $\overline{\text{MS}}$ scheme) has a very well behaved perturbative expansion $\sim 1 + \sum_{i=0} d_i a_s^{i+1}$ in $a_s = \hat{\alpha}_s(M_Z)/\pi$ (see the Appendix for details). The process of analytical continuation from the Euclidean to the physical region induces further terms which are proportional to $\beta$-function coefficients, enhanced by powers of $\pi^2$, and start at $\mathcal{O}(\alpha_s^3)$. Fortunately, these terms [25] involve only known coefficients up to $\mathcal{O}(\alpha_s^5)$, and the only unknown coefficient in $\mathcal{O}(\alpha_s^6)$ is proportional to the four-loop Adler function coefficient, $d_3$. In the massless approximation,

$$\begin{aligned}
\delta_{\text{QCD}}^{\text{NS}} &\approx a_s + 1.4092 a_s^2 - (0.681 + 12.086) a_s^3 + \\
&\quad (d_3 - 89.19) a_s^4 + (d_4 + 79.7) a_s^5 + \\
&\quad (d_5 - 121 d_3 + 3316) a_s^6, \quad (7)
\end{aligned}$$

and terms of order $a_s^7 \sim 10^{-10}$ are clearly negligible. Notice, that the $\mathcal{O}(\alpha_s^6)$ term effectively *reduces* the sensitivity to $d_3$ by about 18%. Eq. (7) amounts to a reorganization of the perturbative series in terms of the $d_i$ times some function of $\alpha_s$; a similar idea is routinely applied to the perturbative QCD contribution to $\tau$ decays [26].

Final state fermion mass effects [22, 27] of $\mathcal{O}(m_f^2)$ (and $\mathcal{O}(m_b^4)$ for $b$ quarks) are best evaluated by expanding in $\hat{m}_q^2(M_Z)$ thus avoiding large logarithms in the quark masses. The singlet contribution of $\mathcal{O}(\alpha_s^2 m_b^2)$ is also included.

The dominant theoretical uncertainty in the $Z$ lineshape determination of $\alpha_s$ originates from the massless quark contribution, and amounts to about $\pm 0.0004$ as estimated in the Appendix. There are several further uncertainties, all of $\mathcal{O}(10^{-4})$: from the $\mathcal{O}(\alpha_s^4)$ heavy top quark contribution to the axial-vector part of $\delta_{\text{QCD}}^{\text{S}}$; from the missing $\mathcal{O}(\alpha\alpha_s^2 m_t^2)$ and $\mathcal{O}(\alpha^2 m_t^2)$ contributions to the $Zb\bar{b}$-vertex; from further non-enhanced but cohering $\mathcal{O}(\alpha\alpha_s^2)$-vertex corrections; and from possible contributions of non-perturbative origin. The total theory uncertainty is therefore,

$$\Delta\alpha_s(M_Z) = \pm 0.0005, \quad (8)$$

which can be neglected compared to the current experimental error. If $\hat{m}_b$ is kept fixed in a fit, then its parametric error would add an uncertainty of $\pm 0.0002$, but this would not change the total uncertainty (8).

Polarization asymmetries are (in some cases up to a trivial factor 3/4 or a sign) given by the asymmetry parameters,

$$A_f = \frac{1 - 4|Q_f|\text{Re}(\hat{\kappa}_f)\hat{s}^2}{1 - 4|Q_f|\text{Re}(\hat{\kappa}_f)\hat{s}^2 + 8Q_f^2 \hat{s}^4 |\hat{\kappa}_f|^2}, \quad (9)$$

and the forward-backward asymmetries by,

$$A_{FB}(f) = \frac{3}{4} A_e A_f. \quad (10)$$

The hadronic charge asymmetry, $Q_{FB}$, is the linear combination,

$$Q_{FB} = \sum_{q=d,s,b} R_q A_{FB}(q) - \sum_{q=u,c} R_q A_{FB}(q), \quad (11)$$

and the hadronic peak cross section, $\sigma_{\text{had}}$, is stored in sigmah, and defined by,

$$\sigma_{\text{had}} = \frac{12\pi \Gamma_{e^+e^-} \Gamma_{\text{had}}}{M_Z^2 \Gamma_Z^2}. \quad (12)$$

Widths and asymmetries are stored in the fields gamma(f), alr(f), and afb(f). The fermion index, f, and the partial width ratios, R(f), are defined in Table 1.

Table 1
Some of the variables used in `lep100.f`. $\Gamma_{\text{inv}}$ and $\Gamma_{\text{had}}$ are the invisible and hadronic decays widths, respectively.

| 0 | $\nu$ | `gamma(0)` $= \Gamma_{\text{inv}}$ | `alr(0)` $= 1$ |
|---|---|---|---|
| 1 | $e$ | `R(1)` $= \Gamma_{\text{had}}/\Gamma_{e^+e^-}$ | `alr(1)` $= A_e$ |
| 2 | $\mu$ | `R(2)` $= \Gamma_{\text{had}}/\Gamma_{\mu^+\mu^-}$ | `alr(2)` $= A_\mu$ |
| 3 | $\tau$ | `R(3)` $= \Gamma_{\text{had}}/\Gamma_{\tau^+\tau^-}$ | `alr(3)` $= A_\tau$ |
| 4 | $u$ | `R(4)` $= \Gamma_{u\bar{u}}/\Gamma_{\text{had}}$ | — |
| 5 | $c$ | `R(5)` $= \Gamma_{c\bar{c}}/\Gamma_{\text{had}}$ | `alr(5)` $= A_c$ |
| 6 | $t$ | `R(6)` $= 0$ | — |
| 7 | $d$ | `R(7)` $= \Gamma_{d\bar{d}}/\Gamma_{\text{had}}$ | — |
| 8 | $s$ | `R(8)` $= \Gamma_{s\bar{s}}/\Gamma_{\text{had}}$ | `alr(8)` $= A_s$ |
| 9 | $b$ | `R(9)` $= \Gamma_{b\bar{b}}/\Gamma_{\text{had}}$ | `alr(9)` $= A_b$ |
| 10 | had | `gamma(10)` $= \Gamma_{\text{had}}$ | `afb(10)` $= Q_{FB}$ |
| 11 | all | `gamma(11)` $= \Gamma_Z$ | — |

**2.4. Fermion masses**

I use $\overline{\text{MS}}$ masses as far as QCD is concerned, but retain on-shell masses for QED since renormalon effects are unimportant in this case. This results in a hybrid definition for quarks. Accordingly, the RG running of the masses to scales $\mu \neq \hat{m}_q$ uses pure QCD anomalous dimensions. The running masses correspond to the functions `msrun(`$\mu$`)`, `mcrun(`$\mu$`)`, etc. which are calculated in `masses.f` to three-loop order. Anomalous dimensions are also available at four-loop order [28], but can safely be neglected. Also needed is the RG evolution of $\alpha_s$ which is implemented to four-loop precision [29] in `alfas.f`.

I avoid pole masses for the five light quarks throughout. Due to renormalon effects, these can be determined only up to $\mathcal{O}(\Lambda_{\text{QCD}})$ and would therefore induce an irreducible uncertainty of about 0.5 GeV. In fact, perturbative expansions involving the pole mass show unsatisfactory convergence. In contrast, the $\overline{\text{MS}}$ mass is a short distance mass which can, in principle, be determined to arbitrary precision, and perturbative expansions are well behaved with coefficients of order unity (times group theoretical factors which grow only geometrically). Note, however, that the coefficients of expansions involving large powers of the mass, $\hat{m}^n$, are rather expected to be of $\mathcal{O}(n)$. This applies, e.g., to decays of heavy quarks ($n = 5$) and to higher orders in light quark mass expansions.

The top quark pole mass enters the analysis when the results on $m_t$ from on-shell produced top quarks at the Tevatron are included. In subroutine `polemasses(nf,mpole)` $\hat{m}_q(\hat{m}_q)$ is converted to the quark pole mass, `mpole`, using the two-loop perturbative relation from Ref. [30]. The exact three-loop result [31] has been approximated (for $m_t$) by employing the BLM [32] scale for the conversion. Since the pole mass is involved it is not surprising that the coefficients are growing rapidly. The third order contribution is 31%, 75%, and 145% of the second order for $m_t$ (`nf` $= 6$), $m_b$ (`nf` $= 5$), and $m_c$ (`nf` $= 4$), respectively. I take the three-loop contribution to the top quark pole mass of about 0.5 GeV as the theoretical uncertainty, but this is currently negligible relative to the experimental error. At a high energy lepton collider it will be possible to extract the $\overline{\text{MS}}$ top quark mass directly and to abandon quark pole masses altogether.

**2.5. $\nu$ scattering**

The ratios of neutral-to-charged current cross sections,

$$R_\nu = \frac{\sigma_{\nu N}^{NC}}{\sigma_{\nu N}^{CC}}, \qquad R_{\bar{\nu}} = \frac{\sigma_{\bar{\nu} N}^{NC}}{\sigma_{\bar{\nu} N}^{CC}}, \qquad (13)$$

have been measured precisely in deep inelastic $\nu$ ($\bar{\nu}$) hadron scattering (DIS) at CERN (CDHS and CHARM) and Fermilab (CCFR). The most precise result was obtained by the NuTeV Collaboration at Fermilab who determined the Paschos-Wolfenstein ratio,

$$R^- = \frac{\sigma_{\nu N}^{NC} - \sigma_{\bar{\nu} N}^{NC}}{\sigma_{\nu N}^{CC} - \sigma_{\bar{\nu} N}^{CC}} \sim R_\nu - rR_{\bar{\nu}}, \qquad (14)$$

with $r = \sigma_{\bar{\nu} N}^{CC}/\sigma_{\nu N}^{CC}$. Results on $R_\nu$ are frequently quoted in terms of the on-shell weak mixing angle (or $M_W$) as this incidentally gives a fair description of the dependences on $m_t$ and $M_H$. One can write approximately,

$$R_\nu = g_L^2 + g_R^2 r, \quad R_{\bar{\nu}} = g_L^2 + \frac{g_R^2}{r}, \quad R^- = g_L^2 - g_R^2, \quad (15)$$

where,

$$g_L^2 = \frac{1}{2} - \sin^2\theta_W + \frac{5}{9}\sin^4\theta_W, \quad g_R^2 = \frac{5}{9}\sin^4\theta_W. \quad (16)$$

However, the study of new physics requires the implementation of the actual linear combinations of effective four-Fermi operator coefficients, $\epsilon_{L,R}(u)$ and $\epsilon_{L,R}(d)$, which have been measured. With the appropriate value for the average momentum transfer, q2, as input, these are computed in the subroutines `nuh(q2,epsu_L,epsd_L,epsu_R,epsd_R)` (according to Ref. [33]), `nuhnutev`, `nuhccfr`, and `nuhcdhs`, all contained in file `dis.f`. Note, that the CHARM results have been adjusted to CDHS conditions [34]. While the experimental correlations between the various DIS experiments are believed to be negligible, large correlations are introduced by the physics model through charm mass threshold effects, quark sea effects, radiative corrections, etc. I constructed the matrix of correlation coefficients using the analysis in Ref. [ 34],

$$
\begin{matrix}
R^- & R_\nu & R_\nu & R_\nu & R_{\bar{\nu}} & R_{\bar{\nu}} & R_{\bar{\nu}}
\end{matrix}
$$

$$
\begin{pmatrix}
1.00 & 0.10 & 0.10 & 0.10 & 0.00 & 0.00 & 0.00 \\
0.10 & 1.00 & 0.40 & 0.40 & 0.10 & 0.10 & 0.10 \\
0.10 & 0.40 & 1.00 & 0.40 & 0.10 & 0.10 & 0.10 \\
0.10 & 0.40 & 0.40 & 1.00 & 0.10 & 0.10 & 0.10 \\
0.00 & 0.10 & 0.10 & 0.10 & 1.00 & 0.15 & 0.15 \\
0.00 & 0.10 & 0.10 & 0.10 & 0.15 & 1.00 & 0.15 \\
0.00 & 0.10 & 0.10 & 0.10 & 0.15 & 0.15 & 1.00
\end{pmatrix}. \quad (17)
$$

The effective vector and axial-vector couplings, $g_V^{\nu e}$ and $g_A^{\nu e}$, from elastic $\nu e$ scattering are calculated in subroutine `nue(q2,gvnue,ganue)` in file `nue.f`. The momentum transfer, `q2`, is currently set to zero [ 36]. Needed is the low energy $\rho$ parameter, `rhonc`, which describes radiative corrections to the neutral-to-charged current interaction strengths. Together with `sin2t0` (described below) it is computed in file `lowenergy.f`.

### 2.6. Low energy observables

The weak atomic charge, `Qw`, from atomic parity violation and fixed target $ep$ scattering is computed in subroutine `apv(Qw,Z,AA,C1u,C1d,C2u,C2d)` where `Z` and `AA` are, respectively, the atomic number and weight. Also returned are the coefficients from lepton-quark effective four-Fermi interactions which are calculated according to [ 37].

These observables are sensitive to the low energy mixing angle, `sin2t0`, which defines the electroweak counterpart to the fine structure constant and is similar to the one introduced in Ref. [ 7]. There is significant correlation between the hadronic uncertainties from the RG evolutions of $\hat{\alpha}$ and the weak mixing angle. Presently, this correlation is ignored, but with the recent progress in atomic parity violation experiments it should be accounted for in the future.

An additional source of hadronic uncertainty is introduced by $\gamma Z$-box diagrams which are unsuppressed at low energies. At present, this uncertainty can be neglected relative to the experimental precision.

Besides `apv`, the file `pnc.f` contains in addition the subroutine `moller` for the anticipated polarized fixed target Møller scattering experiment at SLAC. Radiative corrections are included following Ref. [ 38].

### 2.7. $b \to s\gamma$

Subroutine `bsgamma` returns the decay ratio,

$$R = \frac{\mathcal{B}(b \to s\gamma)}{\mathcal{B}(b \to ce\nu)}. \quad (18)$$

It is given by [ 39, 40],

$$R = \frac{6\alpha}{\pi} \left| \frac{V_{ts}^* V_{tb}}{V_{cb}} \right|^2 \frac{S}{f(z)} \frac{|\bar{D}|^2 + A/S + \delta_{NP} + \delta_{EW}}{(1+\delta_{NP}^{SL})(1+\delta_{EW}^{SL})}, \quad (19)$$

where $|V_{ts}^* V_{tb}/V_{cb}|^2 = 0.950$ is a combination of Cabbibo-Kobayashi-Maskawa matrix elements and $S$ is the Sudakov factor [ 41]. $\delta_{NP}$ and $\delta_{EW}$ are non-perturbative and NLO electroweak [ 42] corrections, both for the $b \to s\gamma$ and the semileptonic ($b \to ce\nu$) decay rates.

$$\bar{D} = C_7^0 + \frac{\hat{\alpha}_s(\hat{m}_b)}{4\pi}(C_7^1 + V), \quad (20)$$

is called the reduced amplitude for the process $b \to s\gamma$, and is given in terms of the Wilson coefficient $C_7$ at NLO. $C_7$ and the other $C_i$ appearing below are effective Wilson coefficients with NLO RG evolution [ 43] from the weak scale to $\mu = \hat{m}_b$ understood. The NLO matching conditions at the weak scale have been calculated in Ref. [ 44]. $\bar{D}$ includes the virtual gluon corrections,

$$V = r_2 C_2^0 + r_7 C_7^0 + r_8 C_8^0, \quad (21)$$

so that it squares to a positive definite branching fraction. On the other hand, the amplitude for gluon Bremsstrahlung ($b \to s\gamma g$),

$$A = \frac{\hat{\alpha}_s(\hat{m}_b)}{\pi}[C_2^0(C_8^0 f_{28}(1) + C_7^0 f_{27}(1) + C_2^0 f_{22}(1)) +$$

$$C_8^0(C_8^0 f_{88}(\delta) + C_7^0 f_{78}(\delta)) + (C_7^0)^2 f_{77}(\delta)], \quad (22)$$

is added linearly to the cross section. The Wilson coefficient $C_2^0$ is defined as in Ref. [ 45]. It enters only at NLO, is significantly larger than $C_7^0$, and dominates the NLO contributions. The parameter $0 \le \delta \le 1$ in the coefficient functions $f_{ij}$ characterizes the minimum photon energy and has been set to $\delta = 0.9$ [ 41], except for the first line in Eq. (22) where $\delta = 1.0$ corresponding to the full cross section. The $f_{2i}$ are complicated integrals which can be solved in terms of polylogarithms up to 5th order. In the code I use an expansion in $z = m_c^2/m_b^2$ and $\delta = 1.0$. Once experiments become more precise the correction to $\delta = 0.9$ should be included.

$f(z)$ is the phase space factor for the semileptonic decay rate including NLO corrections [ 46]. I defined the $\overline{MS}$ mass ratio in $z = [\hat{m}_c(\hat{m}_b)/\hat{m}_b(\hat{m}_b)]^2$ at the common scale, $\mu = \hat{m}_b$, which I also assumed for the factor $\hat{m}_b^5$ multiplying the decay widths. Since I do not re-expand the denominator this effects the phase space function at higher orders. Using the $\mathcal{O}(\alpha_s^2)$-estimate[2]

---
[2]I computed the $\mathcal{O}(\alpha_s^2)$ coefficient for comparison only, and did not include it in the code.

from Ref. [47], I obtain for the semileptonic decay width,

$$\Gamma_{SL} \sim \hat{m}_b^5 f_0(z) \left[ 1 + 2.7 \frac{\hat{\alpha}_s(\hat{m}_b)}{\pi} - 1.6 \left( \frac{\hat{\alpha}_s}{\pi} \right)^2 \right], \quad (23)$$

where $f_0(z)$ is the leading order phase space factor. It is amusing that the coefficients in Eq. (23) are comfortably (and perhaps somewhat fortuitously) small, with the $\mathcal{O}(a_s^2)$-coefficient even smaller than the one in Ref. [47] where a low scale running mass had been advocated. Moreover, using the pre-factor $\hat{m}_b^5$ in the numerator of $R$ reduces the size of $r_7$ in Eq. (21) and therefore the coefficient $\kappa(\delta) = f_{77}(\delta) + r_7/2$ which multiplies the term $a_s(C_7^{0,\text{eff}})^2$. I obtain $-2.1 < \kappa(\delta) < 1.4$, while with the pole mass pre-factor $M_b^5$ one would have $-8.7 \leq \kappa(\delta) < -5.3$.

**Acknowledgements**

It is a pleasure to thank Paul Langacker and Damien Pierce for collaborations on precision analyses. I am grateful to Francesca Borzumati and Paolo Gambino for providing me with parts of their FORTRAN codes.## A. Uncertainties from perturbative QCD

Writing the perturbative expansion of some quantity in its general form for an arbitrary gauge group, it can easily be decomposed into separately gauge invariant parts. Table 2 shows for some (related) examples that after removing the group theoretical pre-factors, all coefficients, $y_i$, are *strictly* of order unity, and that their mean, $\bar{y}$, is very close to zero. In particular, there is no sign of factorial growth of coefficients. These observations offer a valuable tool to estimate the uncertainties associated with the truncation of the loop expansion, so I would like to make them more precise.

Assume (for simplicity) that the $y_i$ are random draws from some normal distribution with unknown mean, $\mu$, and variance, $\sigma^2$. One can show that the marginal distribution of $\mu$ follows a Student-t distribution with $n-1$ degrees of freedom, $t_{n-1}$, centered about $\bar{y}$, and with standard deviation,

$$\Delta \mu = \sqrt{\frac{\sum_i (y_i - \bar{y})^2}{n(n-3)}}. \quad (24)$$

As can be seen from the Table, $\mu$ is consistent with zero in all cases, justifying the nullification of the unknown coefficients from higher loops. I next assert that the distribution of $\sigma$, conditional on $\mu = 0$, follows a scaled inverse-$\chi^2$ distribution with $n$ degrees of freedom, from which I obtain the estimate,

$$\sigma = \sigma_0 \pm \Delta\sigma = \sqrt{\frac{\sum_i y_i^2}{n-2}} \left[ 1 \pm \sqrt{\frac{1}{2(n-4)}} \right]. \quad (25)$$

Inspection of the Table shows indeed that $\sigma$, as the typical size of a coefficient, is estimated to be $\lesssim \mathcal{O}(1)$.

I now focus on the partial hadronic $Z$ decay width. As discussed in Section 2.3, the $\mathcal{O}(\alpha_s^3)$ term, $d_2$, is much smaller than the $\pi^2$ term arising from analytical continuation. This is specifically true for the relevant case of $n_f = 5$ active flavors, where large cancellations occur between gluonic and fermionic loops. Notice, that the $D$-function, in contrast to $R_{\text{had}}$, has opposite signs in the leading terms proportional to $C_A^2 C_F$ and $C_A C_F T_F n_F$. Indeed, the Adler $D$-function and the $\beta$-function have similar structures regarding the signs and sizes of the various terms (see Table 2), and we *do expect* large cancellations in the $\beta$-function. The reason is that it has to vanish identically in the case of $N = 4$ supersymmetry. Ignoring scalar contributions this case can be mimicked by setting $T_F n_f = 2C_A$ (there are 2 Dirac fermions in the $N = 4$ gauge multiplet) or $n_f = 12$ for QCD, which is of the right order. In fact, all known QCD $\beta$-function coefficients become very small for some value of $n_f$ between 6 and 16. We therefore have a reason to expect that similar cancellations will reoccur in the $d_i$ at higher orders. As a $1\sigma$ error estimate for $d_3$, I suggest to use the largest known coefficient $(3 \times 0.71)$ times the largest group theoretical pre-factor in the next order $(C_A^3 C_F)$ which results in

$$d_3 = 0 \pm 77. \quad (26)$$

With Eq. (7) and $\hat{\alpha}_s(M_Z) = 0.120$ one can absorb all higher order effects into the $\mathcal{O}(\alpha_s^4)$-coefficient of $R_{\text{had}}$, $r_3^{\text{eff}} = -81 \pm 63$. This shifts the extracted $\alpha_s$ from the $Z$ line-shape by $+0.0005$ and introduces the small uncertainty of $\pm 0.0004$.

The argument given above does certainly not apply to the *quenched* case, $n_f = 0$, and indeed $d_2(n_f = 0)$ is about $-73\%$ of the $\pi^2$ term, i.e., large and positive. In the case of $n_f = 3$, which is of interest for the precision determination of $\alpha_s$ from $\tau$ decays, $d_2$ is about $-38\%$ of the $\pi^2$ term. If one assumes that the same is true of $d_3$, one would obtain $d_3(n_f = 3) = 60$. Estimates based on the principles of minimal sensitivity, PMS, or fastest apparent convergence, FAC, yield $d_3(n_f = 3) = 27.5$ [25] so there might be some indications for a positive $d_3(n_f = 3)$. In any case, all these estimates lie within the uncertainty in Eq. (26) and we will have to await the proper calculation of the $\mathcal{O}(\alpha_s^4)$-coefficient to test these hypotheses. Note, that the current $\tau$ decay analysis by the ALEPH Collaboration uses $d_3 = 50 \pm 50$ [50] which is more optimistic.

The analogous error estimate for the five-loop $\beta$-function coefficient yields,

$$\beta_4 = 0 \pm 579. \qquad (27)$$

To get an estimate for the uncertainty in the RG running of $\hat{\alpha}_s$, I translate Eq. (27) into

$$\beta_3 = \beta_3 \pm \frac{\hat{\alpha}_s(\mu_0)}{\pi}\beta_4, \qquad (28)$$

where $\mu_0$ is taken to be the lowest scale involved. This overestimates the uncertainty from $\beta_4$, thereby compensating for other neglected terms of $\mathcal{O}(\alpha_s^{n+4} \ln^n \mu^2/\mu_0^2)$. For the RG evolution from $\mu = m_\tau$ to $\mu = M_Z$ this yields an uncertainty of $\Delta\alpha_s(M_Z) = \pm 0.0005$. Conversely, for fixed $\alpha_s(M_Z) = 0.120$, I obtain $\hat{\alpha}_s(\hat{m}_b) = 0.2313 \pm 0.0006$, $\hat{\alpha}_s(m_\tau) = 0.3355 \pm 0.0045$, and $\hat{\alpha}_s(\hat{m}_c) = 0.403 \pm 0.011$, where I have used $\hat{m}_b = 4.24$ GeV and $\hat{m}_c = 1.31$ GeV. For comparison, the ALEPH Collaboration quotes an evolution error of $\Delta\alpha_s(M_Z) = \pm 0.0010$ which is twice as large. I emphasize that it is important to adhere to consistent standards when errors are estimated. This is especially true in the context of a global analysis where the precisions of the observables enter as their relative weights.

Table 2
Coefficients ($\overline{\text{MS}}$) appearing in the $\beta$-function of a simple group [29]; in non-Abelian corrections to the QED $\beta$-function (denoted $\tilde{D}$) [3, 48]; in the Adler $D$-function [49] (rescaled by an overall factor $1/3$); and in $R_{\text{had}}$ (analytical continuation of $D$). The first four segments correspond, respectively, to the first four loop orders of non-singlet type. The fifth segment is the singlet (double triangle) contribution in $\mathcal{O}(\alpha_s^3)$. In $\tilde{D}$, $D$, and $R_{\text{had}}$, an overall factor $\hat{\alpha}M_Z$ and the sums involving charges or $Z$ couplings have been dropped. The completely symmetrical tensors of rank four, $d_A$ and $d_F$, as well as $T_F$ when appearing in parenthesis, apply to the $\beta$-function only. Each $T_F$ is understood to be multiplied by the number of flavors, $n_f$, except for the singlet term involving the symmetrical structure constants, $d$.

| group factor | $\beta$ | $\tilde{D}$ | $D$ | $R_{\text{had}}$ |
|---|---|---|---|---|
| $C_A$ | 0.92 | — | — | — |
| $(T_F)$ | −0.33 | −0.33 | 0.33 | 0.33 |
| $C_A^2$ | 0.71 | — | — | — |
| $C_A T_F$ | −0.42 | — | — | — |
| $C_F(T_F)$ | −0.25 | −0.25 | 0.25 | 0.25 |
| $C_A^3$ | 0.83 | — | — | — |
| $C_A^2 T_F$ | −0.82 | — | — | — |
| $C_A C_F(T_F)$ | −0.36 | −0.23 | 0.18 | 0.18 |
| $C_A T_F^2$ | 0.09 | — | — | — |
| $C_F^2(T_F)$ | 0.03 | 0.03 | −0.03 | −0.03 |
| $C_F T_F(T_F)$ | 0.08 | 0.08 | −0.06 | −0.06 |
| $C_A^4$ | 1.19 | — | — | — |
| $C_A^3 T_F$ | −1.67 | — | — | — |
| $C_A^2 C_F(T_F)$ | −0.23 | −0.28 | 0.32 | −0.38 |
| $C_A^2 T_F^2$ | 0.50 | — | — | — |
| $C_A C_F^2(T_F)$ | −0.42 | −0.07 | 0.51 | 0.51 |
| $C_A C_F T_F(T_F)$ | 0.51 | 0.42 | −0.71 | −0.21 |
| $C_A T_F^3$ | 0.01 | — | — | — |
| $C_F^3(T_F)$ | 0.18 | 0.18 | −0.18 | −0.18 |
| $C_F^2 T_F(T_F)$ | −0.17 | −0.17 | 0.02 | 0.02 |
| $C_F T_F^2(T_F)$ | 0.02 | 0.02 | 0.09 | −0.01 |
| $d_A^2/N_A$ | 1.07 | — | — | — |
| $d_A d_F/N_A$ | −2.38 | — | — | — |
| $d_F^2/N_A$ $(T_F^2 d^2/4)$ | 0.50 | 0.50 | −0.50 | −0.50 |
| $\bar{y}$ | −0.02 | −0.01 | 0.02 | −0.01 |
| $\Delta\mu$ | 0.17 | 0.09 | 0.11 | 0.09 |
| $\sigma_0$ | 0.83 | 0.28 | 0.37 | 0.31 |
| $\Delta\sigma$ | 0.13 | 0.07 | 0.09 | 0.08 |